\documentclass[a4paper,accepted=2023-02-05]{quantumarticle}
\pdfoutput=1

\usepackage[numbers]{natbib}
\usepackage{tabularx}
\usepackage{graphicx}
\usepackage{amsmath}
\usepackage{amssymb}
\usepackage{bm}
\usepackage{diagbox}
\usepackage{color}
\usepackage[table]{xcolor}
\usepackage{amsfonts}
\usepackage[T1]{fontenc}
\usepackage[utf8]{inputenc}

\usepackage{comment}
\usepackage{pgfplots,tikz}
\usepackage[colorlinks=true , citecolor=blue,urlcolor=blue]{hyperref}
\usepackage{braket}

\def\kb{{\mathchar'26\mkern-9mu k}}

\usepackage{ulem}

\def\kbar{{\mathchar'26\mkern-9mu k}}

\def\sf#1{\mathsf{#1}}

\begin{document}
\title{Dynamical many-body delocalization transition of a Tonks gas in a quasi-periodic driving potential }

\author{Vincent Vuatelet}
\affiliation{Univ. Lille, CNRS, UMR 8523  --  PhLAM  --  Laboratoire  de Physique  des  Lasers,  Atomes  et  Mol\'ecules,  F-59000  Lille,  France}
\author{Adam Ran\c con}
\affiliation{Univ. Lille, CNRS, UMR 8523  --  PhLAM  --  Laboratoire  de Physique  des  Lasers,  Atomes  et  Mol\'ecules,  F-59000  Lille,  France}

\begin{abstract}
The quantum kicked rotor is well-known for displaying dynamical (Anderson) localization. It has recently been shown that a periodically kicked Tonks gas will always localize and converge to a finite energy steady-state. This steady-state has been described as being effectively thermal with an effective temperature that depends on the parameters of the kick.
Here we study a generalization to a quasi-periodic driving with three frequencies which, without interactions, has a metal-insulator Anderson transition. We show  that a quasi-periodically kicked Tonks gas goes through a dynamical many-body delocalization transition when the kick strength is increased. The localized phase is still described by a low effective temperature, while the delocalized phase corresponds to an infinite-temperature phase, with the temperature increasing linearly in time. At the critical point, the momentum distribution of the Tonks gas displays different scaling at small and large momenta (contrary to the non-interacting case), signaling a breakdown of the one-parameter scaling theory of localization. 
\end{abstract}

\maketitle

\section{Introduction}
The interplay between disorder and quantum effects has been an intense subject of study for many years, both theoretically and experimentally. A famous effect of the presence of disorder  is the possibility of Anderson localization, where one-body wavefunctions are localized in presence of randomly distributed on-site energy \cite{Anderson1958}.
In three dimensions, states with high enough energy are however delocalized, giving rise to the Anderson metal-insulator transition \cite{Evers2008}. 
It is also well-known that Anderson localization physics can be studied in non-interacting driven quantum systems (without disorder) such as the paradigmatic Quantum Kicked Rotor (QKR). Indeed, such systems display ``dynamical localization'', analogous to Anderson localization but in momentum space \cite{Fishman1982}.  Moreover, it has been shown that by adding two incommensurate frequencies to the QKR, the so-called quasi-periodic QKR (QPQKR), one can induce a localization-delocalization transition in the same universality class as the Anderson transition  \cite{Shepelyansky1987,Casati1989}. This has allowed cold atomic experiments for investigating many features of Anderson physics: observation of Anderson transition \cite{Chabe2008,Lemarie2009}, characterization of its critical properties at the transition \cite{Lemarie2010,Lopez2012}, localization at the lower critical dimension \cite{Manai2015}.

The effects of interactions on disordered systems have been studied extensively   \cite{Nandkishore2015,Abanin2019},  and it is now well understood that although interactions may destroy  localization, a strong enough disorder gives rise to many-body localization (MBL). It is a phase of matter where an extensive number of integrals of motion exists, which breaks ergodicity and prevents the system from thermalizing \cite{Serbyn2013,Huse2014}. The same kind of mechanism prevents driven MBL systems from heating indefinitely \cite{Ponte2015a,Ponte2015}.
In the context of the QKR, the effects of interactions on dynamical localization is a fundamental question that has been studied for various toy models \cite{Adachi1988,Lei2009,Keser2016,Rozenbaum2017,Notarnicola2018,Notarnicola2020}, as well as for the kicked Lieb-Liniger model \cite{Qin2017,Rylands2020,Chicireanu2021,Vuatelet2021}, which is more suited to describe experiments of kicked one-dimensional cold atomic gases. With mean-field arguments, it was shown  both on theoretical and numerical grounds that interactions tend to destroy dynamical localization, giving rise to a subdiffusion in momentum space \cite{Shepelyansky1993,Pikovsky2008,Flach2009,Gligoric2011,Cherroret2014,Lellouch2020}. However, in one dimension, mean-field theory breaks down \cite{Cazalilla2011}. Rylands \textit{et al.} have argued that dynamical localization persists in presence of interaction, leading to a Many-Body Dynamically Localized (MBDL) phase \cite{Rylands2020}. An investigation of the kicked Lieb-Liniger model in the infinite interaction limit (Tonks regime) showed that MBDL indeed persists but alters drastically the exponential decay of the momentum distribution due to the interactions \cite{Vuatelet2021}. Furthermore, in that work, it was shown that the MBDL steady-state of a Tonks gas starting from its ground state can be described as being effectively thermal, with an effective temperature and chemical potential depending on the parameters of the kick.

Two recent experiments have studied the fate of dynamical localization for interacting cold gases of bosons \cite{Cao2022,SeeToh2022}, observing a destruction of dynamical localization, in apparent contradiction with \cite{Rylands2020,Vuatelet2021}, and more in line with the mean-field results. The latter might not be too surprising concerning \cite{Cao2022} since it was performed with a three-dimensional gas at weak interactions, for which a mean-field description should be relevant. On the other hand,  \cite{SeeToh2022} used a Bose gas confined in a transverse two-dimensional optical lattice, in the  quasi-one-dimensional regime where MBDL would be expected. We note however that the gas is  weakly interacting whereas numerical evidence, and arguments in the Tonks regime, are valid for strong interactions \cite{Rylands2020,Vuatelet2021}. It is still unclear if there is a true transition between a delocalized regime at weak interaction and MBDL phase at strong interactions, or if a one-dimensional kicked gas always localizes but with a subdiffusive regime at short times and weak coupling. We note that two weakly interacting particles always dynamically localize when kicked \cite{Chicireanu2021}, but large number of particles are extremely difficult to simulate in this regime, so new theoretical methods are necessary. On the experimental side, longer time scales should be studied, the major obstacle right now being the trapping potential along the kick direction, which allows only for a few hundred of kicks.   

In the present manuscript, we focus instead on the interacting QPQKR, which can be experimentally studied as in \cite{Cao2022,SeeToh2022} by a simple modification of the kicking sequence.
In this case, mean-field analyses have found that the localized-diffusion transition is replaced by a subdiffusion-diffusion transition characterized by new critical properties and a two-parameter scaling law \cite{Cherroret2014,Ermann2014,Vermersch2020}. An analysis beyond mean-field is still missing.
Here we study a quasi-periodically kicked Lieb-Liniger gas in the Tonks regime. We show that the system displays a many-body dynamical localization-delocalization transition, between a localized phase with asymptotically finite energy and a delocalized phase that heats up to infinite temperature. We show that the two phases can be described as being effectively thermal in the same line as \cite{Vuatelet2021}, with the delocalized phase having a temperature linearly increasing in time. While the phase diagram is the same as that of the non-interacting system, we show that the critical state of the Tonks gas is markedly different from that of free particles.

The article is organized as follows. In Sec.~\ref{sec_model}, we present the model and the physical observables studied, also recalling the physics of non-interacting QPQKR. In Sec.~\ref{sub_sec_loc} and Sec.~\ref{sub_sec_deloc} we describe our numerical results for the momentum distribution and the coherence function in both localized and delocalized regimes respectively. We also give an interpretation of those results in terms of effective thermalization. In Sec.~\ref{sub_sec_crit}, we analyze the critical regime, and the scaling laws obeyed by the momentum distribution. Finally, a discussion of our results is given in Sec.~\ref{conclusion}.

\section{The model\label{sec_model}}

We consider a system of $N$ interacting bosons of mass $m$, described by the following quasi-periodic Hamiltonian
\begin{equation}
\begin{split}
\hat{\mathcal{H}}(t)=&\sum_{i}\left(\frac{\hat p_{i}^{2}}{2}+\mathcal{K}(t)\cos(\hat x_{i})\sum_{n}\delta(t-n)\right)\\&+g\sum_{i<j}\delta(\hat x_{i}-\hat x_{j}),
\end{split}
\label{eq_H}
\end{equation}
where $\mathcal{K}(t)=K(1+\varepsilon\cos(\omega_{1}t)\cos(\omega_{2}t))$, $K$ is the kick strength, $\varepsilon$ is the modulation strength, and the $\omega_{1,2}$ are two incommensurate frequencies with respect to $2\pi$. From now on, they will be fixed to $2\pi\sqrt{5}$ and $2\pi\sqrt{13}$. 
All physical quantities are dimensionless in Eq.~\eqref{eq_H}: time is in units of the period of the kicks $T_1$, length are in units of kick-potential wavenumber $1/k_{K}$, momenta are in units of $m/T_1 k_K$. Thus, position and momentum obey the commutation relations $[\hat{x}_{i},\hat{p}_{j}]=i\kbar\delta_{ij}$ with $\kbar=\hbar k_{K}^{2}T_1/m$ the (dimensionless) effective Planck constant.  The system size $L$ is set equal to $2\pi$, and we assume periodic boundary conditions, which imply that momenta are quantized in units of $\kbar$ (we also take the Boltzmann constant $k_{B}=1$).


\begin{figure}[t!!!!!!!!!!!!]
\centering
\includegraphics[scale=0.2]{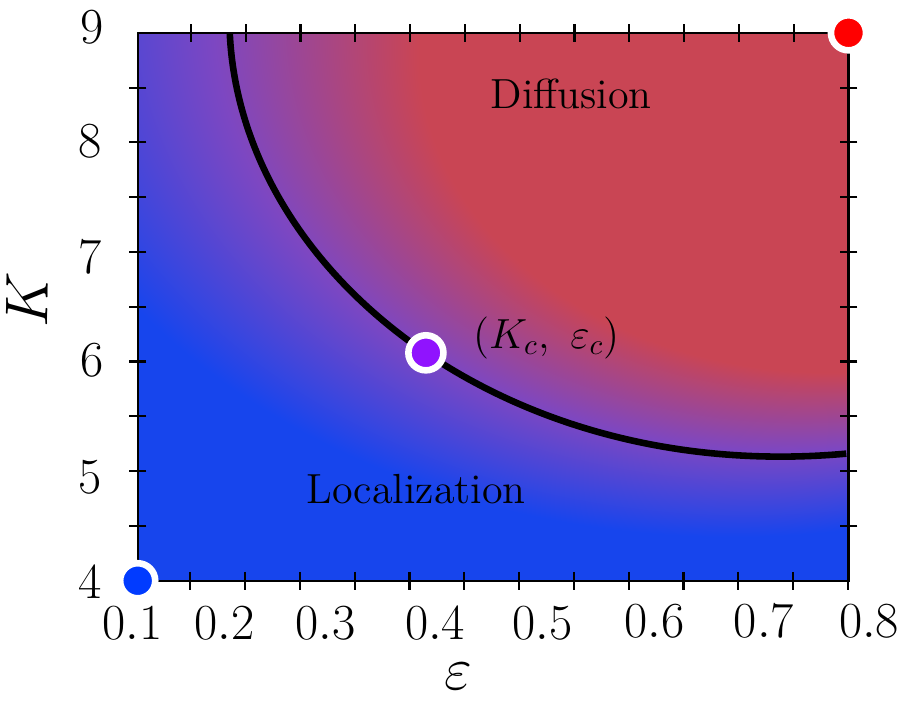}
\includegraphics[scale=0.18]{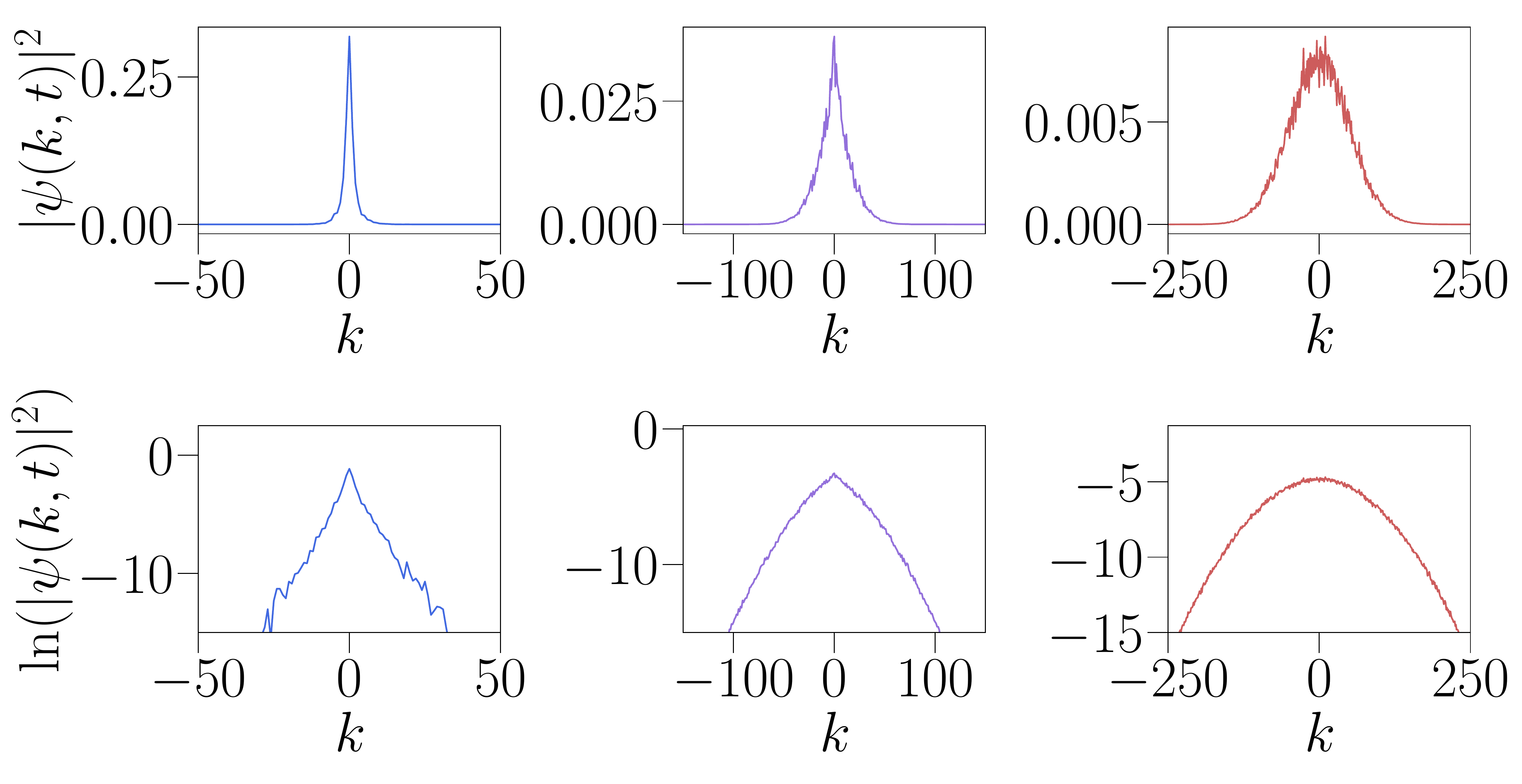}
\caption{Top panel: Schematic representation of the phase diagram of the QPQKR as a function of $K$ and $\varepsilon$ for $\kbar\simeq 2.9$, $\omega_{1}=2\pi\sqrt{5}$ and $\omega_{2}=2\pi\sqrt{13}$. The blue (red) area represents the localized (delocalized) phase. The purple area corresponds to the set of $(K,\varepsilon)$ where the system is critical. Middle and bottom panel: Behavior of the single-particle wave-function, in linear and log scale respectively, in momentum space at $t=500$ kicks and different sets of $(\kbar,K,\varepsilon)$. From left to right: $(\kbar,K,\varepsilon)=(2.89,4.0,0.1)$ (localized), $(\kbar,K_{c},\varepsilon_{c})\simeq(2.85,6.36,0.43)$ (critical) \cite{Lemarie2009}, and $(\kbar,K,\varepsilon)=(2.89,9.0,0.8)$ (delocalized). }
\label{fig:phasediagramepsvsk}
\end{figure}

We start by recalling the properties of the system without interaction ($g=0$) \cite{Lemarie2009}. In this limit, we recover the physics of the QPQKR, which strongly depends on the values of $(K,\varepsilon)$ as  shown in Fig.~\ref{fig:phasediagramepsvsk}. At low values of $(K,\varepsilon)$, the system is localized (see the blue region of Fig.~\ref{fig:phasediagramepsvsk}, top panel) with the single-particle wave-function being exponentially localized with an associated ``localization length'' $p_{loc}$. In particular, the energy saturates at long time, i.e. $E(t) = E_{0}+\frac{p_{loc}^{2}}{2}$, with $E_0$ the initial energy. For large values of $(K,\varepsilon)$, the system's energy increases linearly in time, i.e. $E(t)=E_{0}+2\mathcal{D}t$, $\mathcal{D}$ the diffusion coefficient,  while the  single-particle wave-function is Gaussian (red region in the top panel of Fig.~\ref{fig:phasediagramepsvsk}). There is a critical line separating the two phases, where the energy grows with an anomalous scaling, $E(t)\sim t^{2/3}$. 
Close to criticality, the localization length diverges as $p_{loc}\propto |\delta|^{-\nu}$ in the localized phase, and the diffusion coefficient vanishes as $\mathcal D\propto |\delta|^{\nu}$, where $\delta$ is the distance to the critical point and $\nu\simeq 1.57$ is the associated critical exponent \cite{Slevin1999}. 

Right at criticality, the single-particle wave function is known to be  a function of $k t^{-1/3}$ only, a consequence of the one-parameter scaling theory of localization \cite{Abrahams1979},
\begin{equation}\label{eq:scaling}
    |\psi(k,t)|^{2}=t^{-1/3}f(k t^{-1/3}).
\end{equation}
The scaling function $f$ can be  estimated using a self-consistent theory for the diffusion coefficient, similar to the one developed for disordered systems, see \cite{Wolfle2010} for a review. This self-consistent theory has been shown to capture the Anderson transition qualitatively, and to describe quantitatively the momentum distribution of the critical PSQKR measured experimentally -- at least for large enough momenta.  Its shape is given by \cite{Lemarie2010,Lopez2013}
\begin{equation}\label{eq:scalinglaw}
    |\psi(k,t)|^{2}=\frac{3}{2}\left(3\gamma^{3/2}t\right)^{-1/3}{\rm Ai}\left[\left(3\gamma^{3/2}t\right)^{-1/3}|k|\right],
\end{equation}
where $\gamma=\Gamma(2/3)\Lambda_{c}/3$,  $\Lambda_{c}=\lim_{t\to\infty}\braket{p ^{2}(t)}t^{-2/3}$, $\Gamma$ the Gamma function, and ${\rm Ai}(x)$ the Airy function.
 The self-consistent theory is however unable to capture the strong fluctuations induced by criticality: it does not predict the correct critical exponents and does not capture multifractality. The latter shows up in disordered systems in the scaling of the inverse participation ratios, which do not obey standard scaling relations due to the strong fluctuations of the critical eigenstates, with regions where they are unexpectedly large and regions where they are unexpectedly small \cite{Rodriguez2011}.
In the context of the PSQKR, multifractality modifies the very long time behavior of the momentum distribution at small momenta. It has been shown to behave as  \cite{Panayotis2019}
\begin{equation}\label{eq:multifractal}
    |\psi(k,t)|^{2}=t^{-1/3}(\alpha-\beta|k t^{-1/3}|^{d_2-1}),
\end{equation}
with $\alpha,\beta>0$, and the multifractal dimension $d_2\simeq 1.24$ \cite{Rodriguez2011}. Importantly, these multifractal corrections do not destroy the scaling form Eq.~\eqref{eq:scaling}. There is however a subtlety: the multifractal corrections grow in time and stabilizes only at very long times, of the order of $10^8$ kicks. One can nevertheless fit an effective, time-dependent, multifractal dimension $d_{2,eff}(t)$ at shorter times \cite{Panayotis2019}. The cross-over scale between the Airy shape and the multifractal behavior is given by the mean distance (in momentum space) traveled by the particles in a time $t$, $L_t=l(t/\tau)^{1/3}$, with $l$ the mean-free path, and $\tau$ the mean-free time \cite{Panayotis2019,Chalker1990}.

In this article, we focus on the infinite interaction limit ($g\to\infty$), the so-called Tonks regime. In this regime, one can compute the exact time-evolution of the bosonic many-body wave function $\Psi_{B}(\{x\},t)$ at positions $\{x\}=\{x_1,\ldots,x_N\}$ thanks to the Bose-Fermi mapping \cite{Girardeau1960,Lenard1964,Buljan2008,Jukic2008,Pezer2009}
\begin{equation}
\Psi_{B}(\{x\},t)=\prod_{i<j} {\rm sign}(x_{i}-x_{j})\Psi_{F}(\{x\},t),
\end{equation}
where $\Psi_{F}(\{x\},t)$ is the fermionic wavefunction of $N$ free fermions, that can be simply expressed as the Slater's determinant of single-particle wavefunctions $\psi_i(x)$, evolving according to the Schr{\"o}dinger's equation of the non-interacting QPQKR (i.e. $\hat{\mathcal H}(t)$ with $g=0$). Since the dynamic of each single-particle orbitals $|\psi_{i}(t)\rangle$ is that of one of the non-interacting QPQKR, we can infer that they will all be either localized, delocalized, or critical, depending on the parameters, as discussed above.
We will always assume that the system starts in its ground state, i.e. the fermionic momentum distribution describes a Fermi sea filled up to the Fermi momentum $p_{F}\propto N$, with ground-state energy denoted $E_{0}$.

For a Tonks gas, any local observable, such as the energy or the density, is the same as the fermionic one. 
On the other hand, non-local observables such as the one-body density matrix 
\begin{equation}
\begin{split}
\rho(x,y,t)=N\int dx_{2}\ldots dx_{N}&\Psi_{B}^{*}(x,x_{2},\ldots,x_{N};t)\times\\ &\Psi_{B}(y,x_{2},\ldots,x_{N};t),
\end{split}
\end{equation}
and its Fourier transform, the momentum distribution
\begin{equation}
n_{k}(t)=\frac{1}{L}\int dxdy e^{ik(x-y)}\rho(x,y,t),
\end{equation}
are significantly different from those of free fermions. Since the phenomena we focus on are non-local in real space, we expect those observables to be drastically different from  that at $g=0$. We focus on those observables at long time in each of the three phases in the following.
Finally, we will also study the fermionic momentum distribution, given by
\begin{equation}
n^F_{k}(t)=\sum_i|\langle k| \psi_i(t)\rangle|^2.
\end{equation}

A great advantage of the Tonks limit is that the time-evolution of the many-body wavefunction of $N$ particles can be computed numerically to arbitrary precision by increasing the size of basis in which we write the single particle orbitals. Since in the context of the PSQKR the relevant physics takes place in momentum space, we truncate the wavevector basis $|k\rangle$ using a cut-off $M$ large enough (depending on the phase) such that the single-particle orbitals $\psi_i(k,t)$ are converged (i.e. $|\psi_i(\pm M,t)|\simeq 0$ at all time $t$). Their dynamics is obtained numerically by alternating between real space for kicks and momentum space for free propagation via fast Fourier transform. The system starting in the ground state, we choose $\psi_i(k,0)=\delta_{k,k_i}$, with $k_i$ the $N$ smallest wavevectors. Once the single-particle orbitals are obtained, the bosonic observables are computed straightforwardly using the method of Refs.~\cite{Rigol2005a,Rigol2005}. Noting that the momentum distribution of the Tonks gas decays as a power law at large momenta (see below), we expect that the momentum distribution of the Tonks gas will not be correctly captured for $k\gtrsim M$ due to the finite basis in momentum space. This can be improved by increasing the cut-off.

\begin{figure}[t!!]
\includegraphics[scale=0.175,clip]{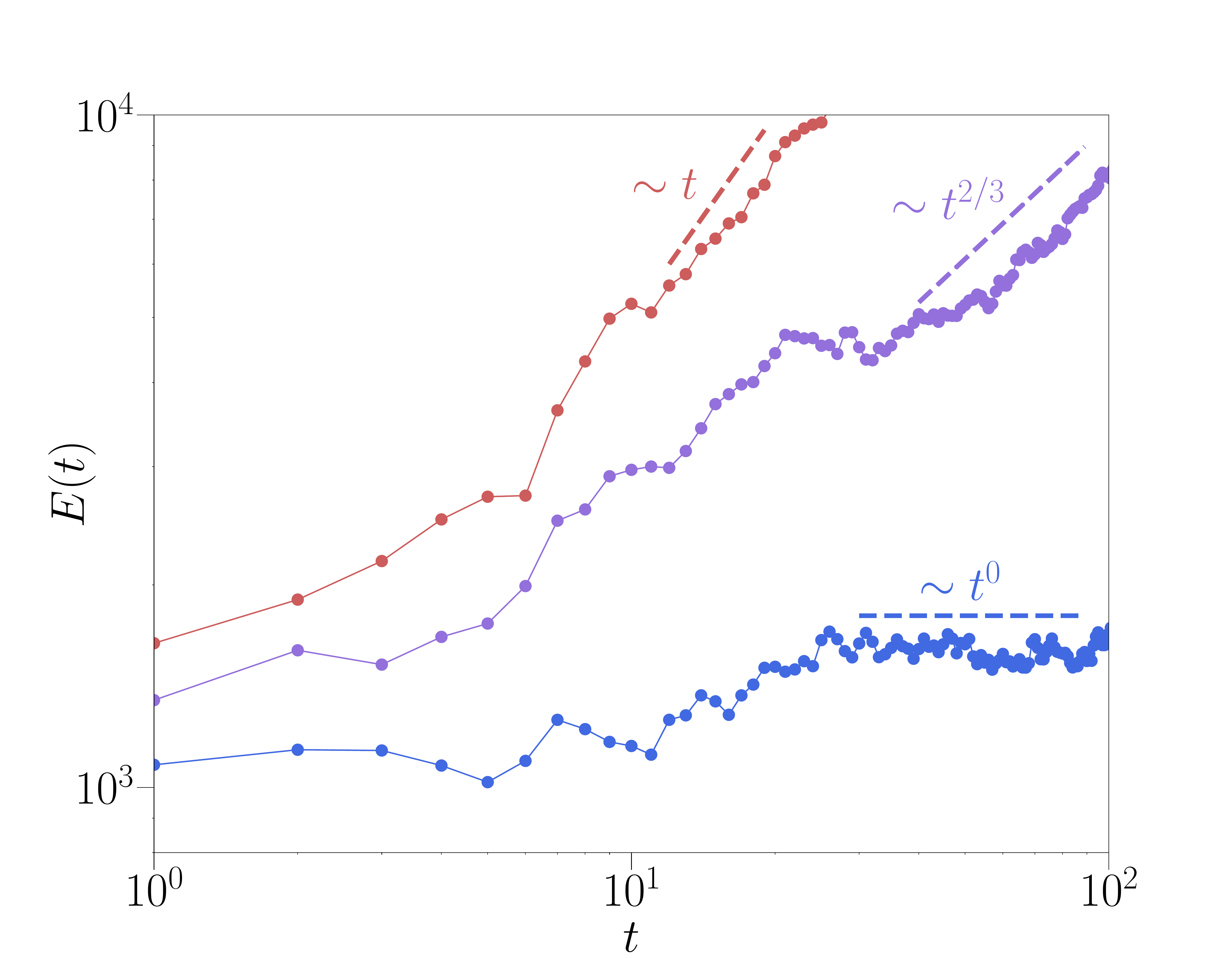} 
\caption{Time evolution of the total energy in log-log scale for  $N=11$ particles in the delocalized $(K=8,\kb=2.89,\varepsilon=0.55)$, 
 critical $(K=6.36,\kb=2.85,\varepsilon=0.43)$ and loacalized $(K=3,\kb=2.89,\varepsilon=0.55)$ regimes (top to bottom). Dashed lines correspond to the expected power law growth of the energy in each regime.}
\label{fig:localized energy}
\end{figure}

\section{Study of the dynamics at long time\label{sec_phase_study}}

The phase diagram of the quasi-periodically kicked Tonks gas can be understood easily. Indeed, since the energy of the Tonks gas is the same as that of free fermions, and since all the single-particle orbitals are either localized, delocalized, or critical, depending on $K$ and $\varepsilon$, the phase diagram is also given by the top panel of Fig.~\ref{fig:phasediagramepsvsk}. The time evolution of the energy in the three cases is shown in Fig.~\ref{fig:localized energy}. We now analyze the three phases in turn.

\subsection{Localized phase\label{sub_sec_loc}}

At small values of $(K,\varepsilon)$, all single-particle orbitals are dynamically localized. Here we use $(\kb,\varepsilon)=(2.89,0.55)$ in our numerical simulations. This case is very similar to that of a (single-frequency) kicked Tonks gas discussed in detail in \cite{Vuatelet2021}. We will therefore only briefly analyze this phase, recalling only the relevant information needed to contrast with the other phases.

\begin{figure}[t!]
\includegraphics[scale=0.2,clip]{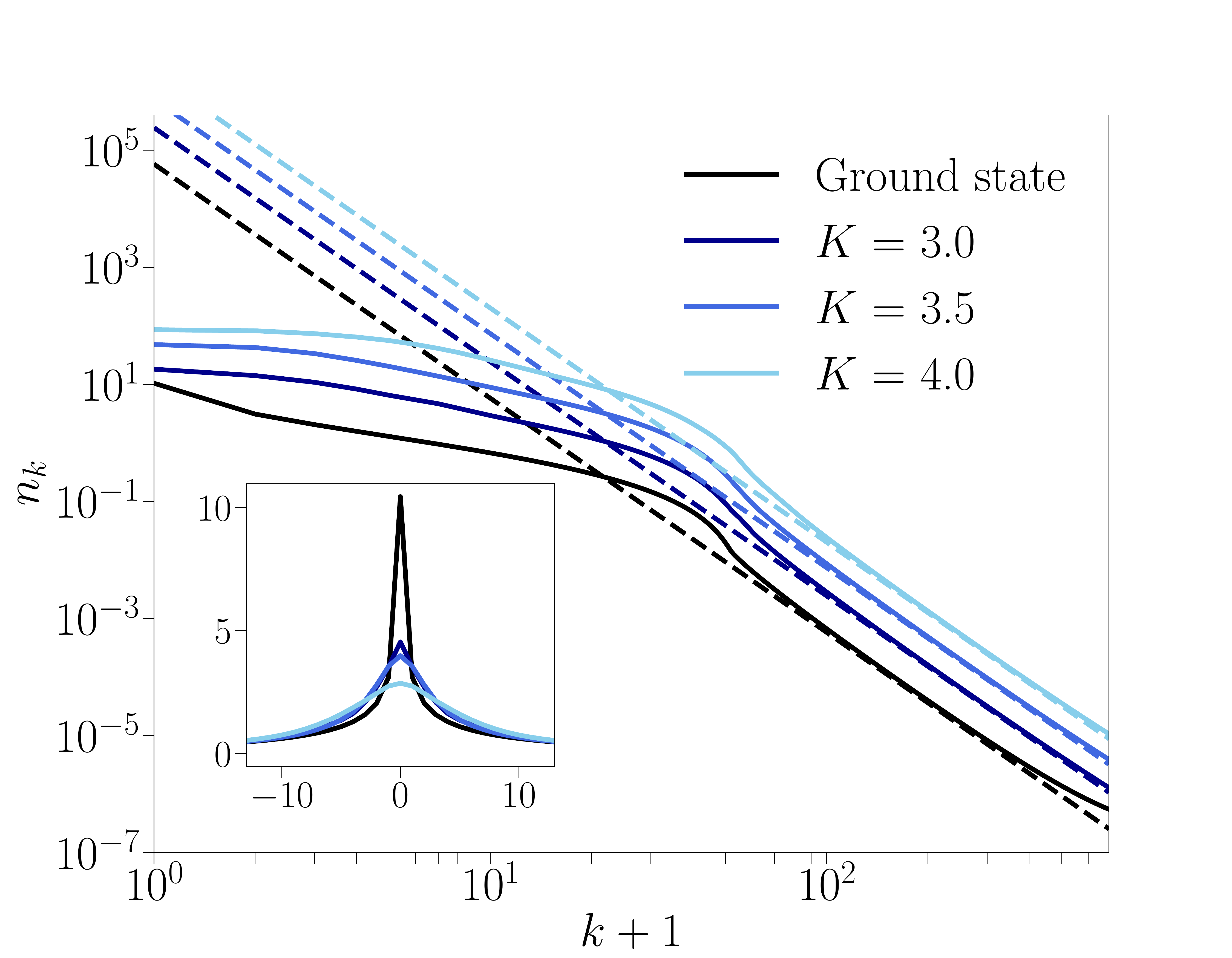} 
\caption{Steady-state momentum distribution  in log-log scale in the localized regime at $t=1000$ kicks for $K = 3$, $3.5$ and $4$, $N = 51$
particles (for better visibility in the main panel, the different curves have been shifted  vertically by adding a constant to $n_k$, from top to bottom: 30, 12, 4, 0).} The dashed line shows the asymptotic behavior
$n_{k}\simeq \mathcal{C}_{ss}/k^{4}$
at large momenta, with $\mathcal{C}_{ss}$ computed using the
effectively thermal density matrix (see text). The inset shows the same quantities in linear scale.

\label{fig:bosonic distribution localized phase}
\end{figure}

First, since the fermions dynamically  localize, their energy saturates at long time to $E=E_{0}+N\frac{p_{loc}^{2}}{2}$, as shown in Fig.~\ref{fig:localized energy} in blue. This implies that the energy of the Tonks gas also saturates and this can be interpreted as Many-Body Dynamical Localization \cite{Rylands2020,Vuatelet2021}.
However, the momentum distribution of the Tonks gas is very different from that of a dynamically localized single-particle. Indeed, at large momenta, the momentum distribution does not decay exponentially, but as a power-law, $n_{k}\sim \mathcal{C}/k^{4}$, where $\mathcal{C}$ is the so-called Tan's contact \cite{Olshanii2003,Tan2008}, see Fig.~\ref{fig:bosonic distribution localized phase}. Furthermore, contrary to the initial state (the ground state), which displays quasi-long-range order and a divergence at small momenta $n_{k}(t=0)\propto k^{-1/2}$ \cite{Cazalilla2011}, this divergence is rounded at long time in the localized phase (see inset).

\begin{figure}[t]
\includegraphics[scale=0.2,clip]{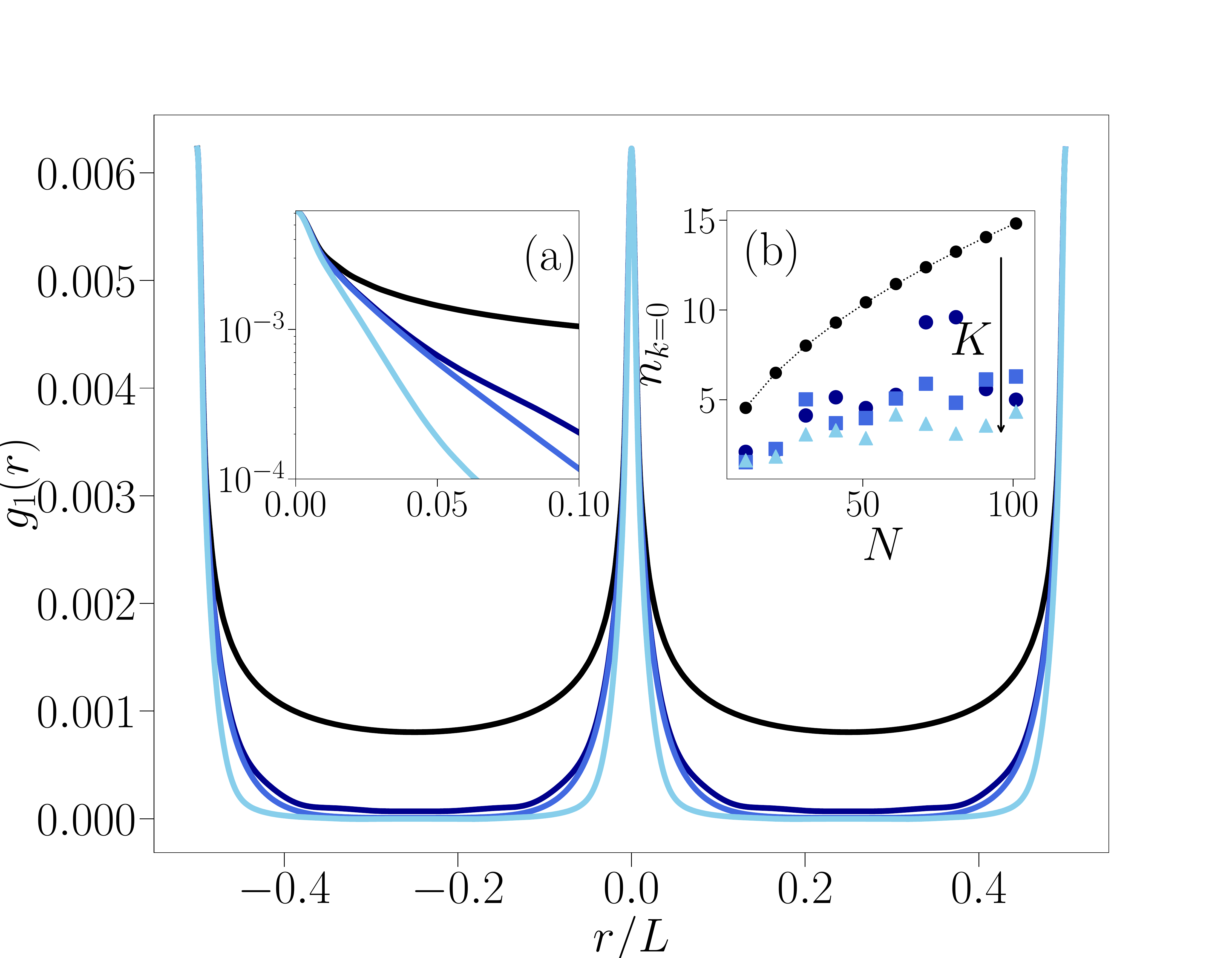} 
\caption{Steady-state coherence function $g_{1}(r)$ with same parameters and legend as Fig.~\ref{fig:bosonic distribution localized phase}. Insets (a) Same
data, in semi-log scale, emphasizing the exponential decay in
the MBDL, compared to the $1/\sqrt{r}$ decay of the initial condition (solid black curve); (b) Occupation of the zero-momentum state $n_{k=0}$. It grows as $\sqrt{N}$ in the ground state (dotted line) but saturates to a finite value in the MBDL regime.}
\label{fig:g1_loc}
\end{figure}

The same information can be gathered by studying the coherence function 
\begin{equation}
g_{1}(r)=\frac{1}{L}\int dR \rho(R-r/2,R+r/2).
\end{equation}
In the ground state, $g_{1}(r,t=0)\sim 1/\sqrt{|r|}$ for $r p_F\gg 1$ due to quasi-long-range order. The localized coherence functions shown in  Fig.~\ref{fig:g1_loc} decay exponentially fast with the distance, $g_{1}(r)\sim e^{-2|r|/r_c}$, implying the absence of coherence. Thus the kicks have destroyed the phase coherence of the gas, which is in agreement with the fact that $n_{k=0}$ no longer scales with the number of particles (see inset (b) of Fig.~\ref{fig:g1_loc}).

We have shown in \cite{Vuatelet2021} that this phenomenology can be described by assuming that the system is effectively thermal, i.e. the steady-state density matrix of the system $\hat\rho_{ss}$ can be written as a thermal density matrix
\begin{equation}
\hat \rho_{ss}\propto e^{-(\hat{\mathcal H}_{TG}-\mu_{eff}\hat{N})/T_{eff}},
\end{equation}
where $\hat{\mathcal H}_{TG}$ is Hamiltonian of the Tonks gas without driving, and  $T_{eff}$ and $\mu_{eff}$ are respectively the effective temperature and chemical potential, that depend on the steady-state.

\begin{figure}[t]
\centering
\includegraphics[scale=0.2,clip]{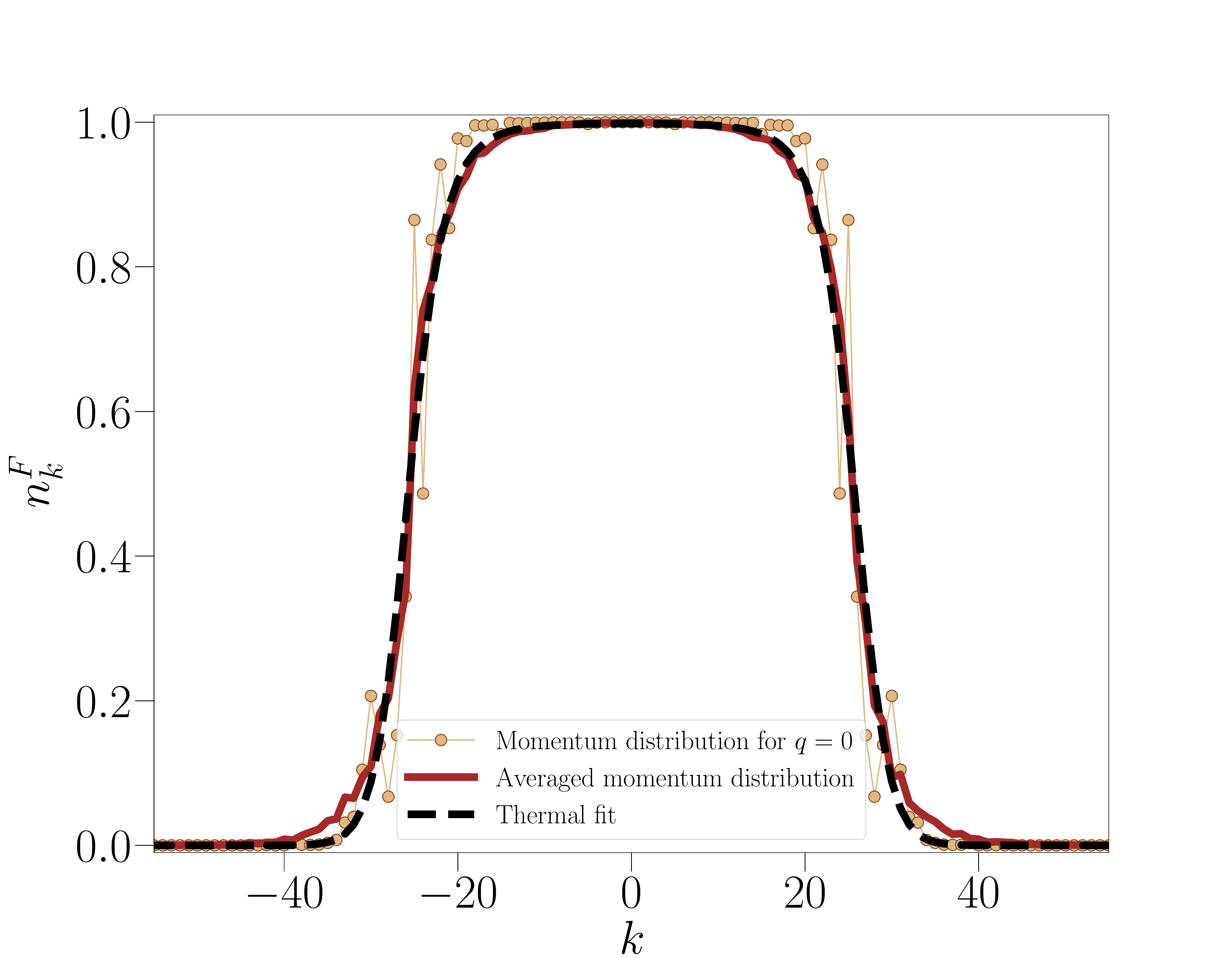}
\caption{Comparison between raw and averaged distribution for the many-body momentum distribution in the localized regime for $N=51$, $K=3.5$, $t=1000$.}
\label{fig:comparaisonaveragedandrawdistributionfermionnf51fitthermal}
\end{figure}

To extract these parameters, it is more convenient to study the steady-state momentum distribution of the fermions, which is however very noisy.
Following the procedure of \cite{Vuatelet2021}, the effective temperature and chemical potential can be obtained by a modified QPQKR Hamiltonian depending on a parameter $q$ (playing the role of a quasi-momentum in the non-interacting QKR)
\begin{equation}
\hat{\mathcal{H}}_{q}=\frac{(\hat{p}+q\kb)^{2}}{2}+\mathcal{K}(t)\cos(\hat{x})\sum_{n}\delta(t-n).
\end{equation}
It allows for smoothing the momentum distribution of the fermions and for an easier extraction of the effective parameters.
In Fig.~\ref{fig:comparaisonaveragedandrawdistributionfermionnf51fitthermal}, in addition to the momentum distribution $n_{k}^{F}$ at $q=0$, we show the momentum distribution averaged over 150 values of $q$ (full line). The smoothing effect of the procedure is clear here. The same figure shows the corresponding Fermi-Dirac at an effective temperature $T_{eff}$ and effective chemical potential $\mu_{eff}$ such that the fitting distribution describes the data. The effective temperature and chemical potential are obtained using imposing that
\begin{equation}\label{eqfFD}\begin{split}
&\sum_{k\in\mathbb{Z}}f_{FD}(k,T_{eff},\mu_{eff})=N,\\
&\sum_{k\in\mathbb{Z}}\frac{\kb^{2}k^{2}}{2}f_{FD}(k,T_{eff},\mu_{eff})=E,
\end{split}
\end{equation}
where $E$ is the energy in the steady state obtained from the averaged momentum distribution, and $f_{FD}$ is the Fermi-Dirac distribution
\begin{eqnarray}
f_{FD}(k,T,\mu)=\frac{1}{e^{\frac{\kb^{2}k^{2}/2-\mu}{T}}+1}.
\end{eqnarray}

\begin{figure}[t!]
\centering
\includegraphics[scale=0.2,clip]{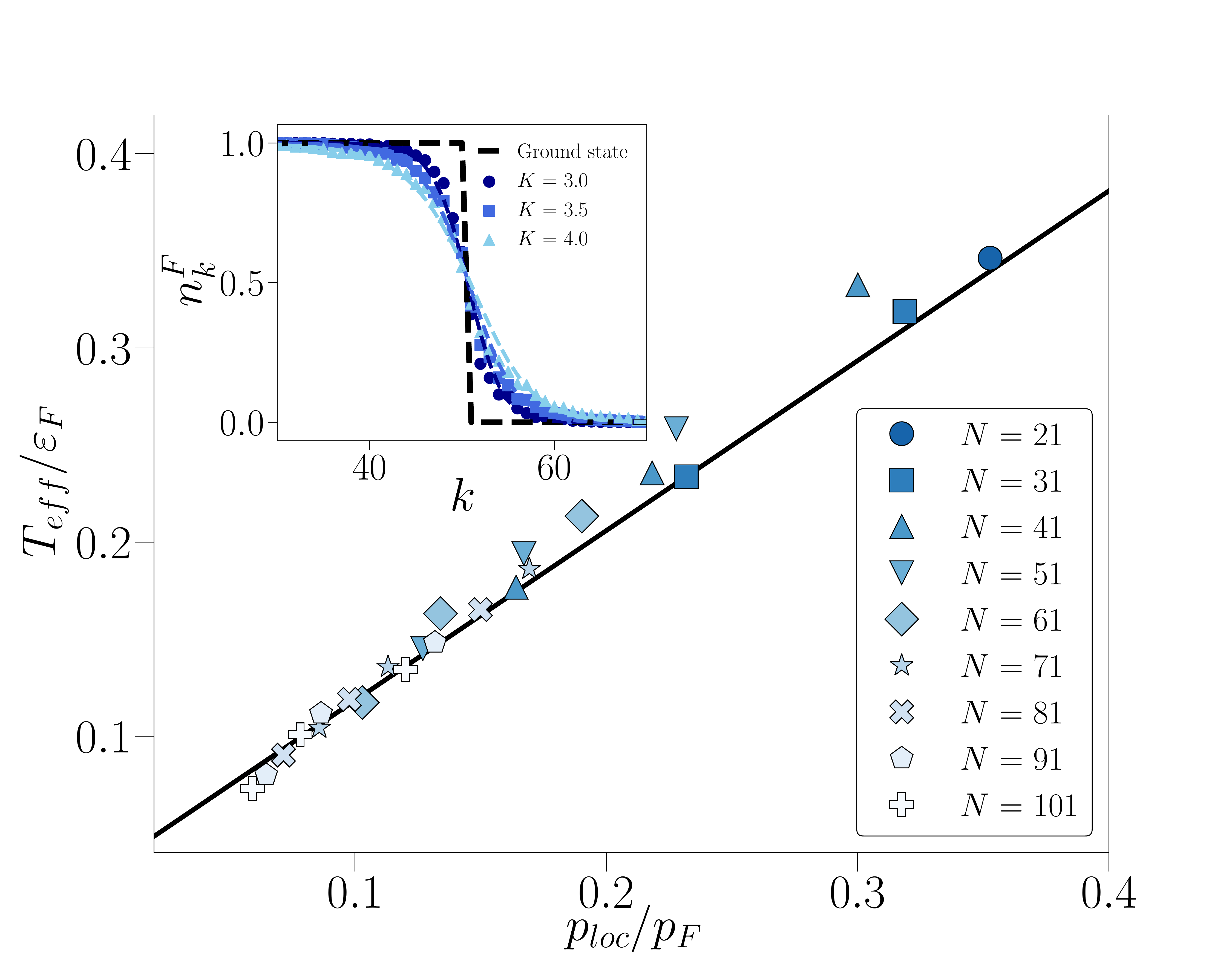}
\caption{Effective temperature $T_{eff}/\varepsilon_F$ as a function of $p_{loc}/p_F$ in the localized regime for
various $N$ and $K$ (corresponding to various $p_F$ and $p_{loc}$, respectively).  The collapse of the data shows the linear
scaling for small enough $p_{loc}/p_F$, $T_{eff}/\varepsilon_F \simeq \frac{2\sqrt{3}}{\pi}
p_{loc}/p_F$ (black line).
Inset: Momentum distribution of the fermions $n^{F}_{k}$ in the localized
regime (symbols), fitted by a Fermi-Dirac distribution with temperature $T_{eff}$ and chemical potential $\mu_{eff}$, for $N=101$, $t=1000$.}
\label{fig:teffepsvsplocoverpf}
\end{figure}

We observe that the fit is very good in the localized phase, see the inset of Fig.~\ref{fig:teffepsvsplocoverpf}. In this phase, $p_{loc}$ is much smaller than $p_{F}$ (due to the small values of $K$). This corresponds to low effective temperatures in comparison to the initial Fermi energy $\varepsilon_{F}=p_{F}^{2}/2$. Thus, we can derive an explicit expression of the effective temperature as a function of $p_{loc}$ and $p_{F}$ by using a Sommerfeld expansion of the energy  \cite{Vuatelet2021}
\begin{equation}
\label{eq:teff_loc}
\frac{T_{eff}}{\varepsilon_{F}}=\frac{2\sqrt{3}}{\pi}\frac{p_{loc}}{p_{F}}.
\end{equation}
Fig.~\ref{fig:teffepsvsplocoverpf} shows that Eq.~(\ref{eq:teff_loc}), deep in the localized phase, describes very well the fitted effective temperatures.\\

This information about the thermal properties of the fermions can be used to describe the localized phase of the Tonks gas since the properties of the thermal Tonks gas are well-understood \cite{Cazalilla2011}.  For a thermal Tonks gas of $N$ bosons at temperature $T$, Tan's contact is given by \cite{Vignolo2013}
\begin{equation}
\mathcal{C}_{th}(N,T)=\frac{8NE(N,T)}{L^{2}\kb^{2}}.
    \label{eq_thermalcontact}
\end{equation}
 In the context of effective thermalization, we expect that the contact in the localized regime $\mathcal{C}$ to be given by $\mathcal{C}=\mathcal{C}_{th}(N,T_{eff})$. Fig.~\ref{fig:bosonic distribution localized phase} shows that the power-law decay is very well described by $\mathcal{C}_{th}(N,T_{eff})/k^{4}$ (dashed lines).

\begin{figure}[t!]
\centering
\includegraphics[scale=0.2,clip]{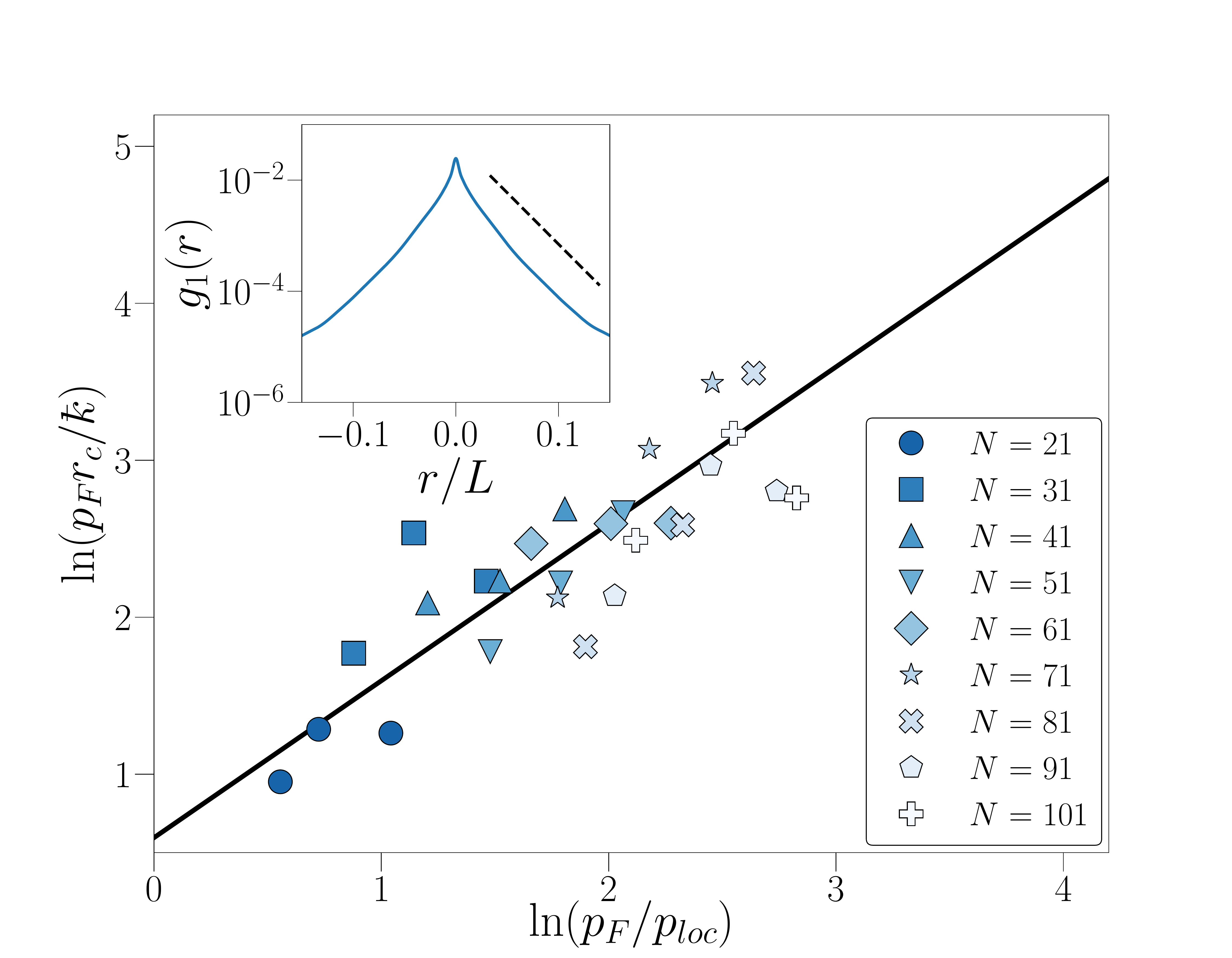}

\caption{Correlation length $r_c$ as a function of $p_{loc}$ in the localized regime for
various $N$ and $K$ (corresponding to various $p_F$ and $p_{loc}$, respectively).
The collapse of the data shows that it is independent of the particle
number. The line corresponds to the scaling $r_{c}=\frac{\pi}{3}\frac{\kb}{p_{loc}}$. Inset: Coherence function $g_1(r)$ (blue line) and the expected exponential decay (dashed line)
with $r_{c}=\frac{\pi}{3}\frac{\kb}{p_{loc}}$, for $N = 101$, $K = 4$, $t=1000$.}
\label{fig:pfrcvspfoverplock3435kbar2}
\end{figure}

The exponential decay of $g_{1}$ in a thermal Tonks gas is known \cite{Cazalilla2011,Its1991}, and in particular at low temperature we have $r_{c}=\frac{\kb v_{F}}{T_{eff}}$, where $v_{F}=\frac{\kb N}{2}$ is the Fermi velocity. From the relation deduced between $T_{eff}$ and $p_{loc}$, we expect $r_{c}$ to be independent of particle and inversely proportional to $p_{loc}$, i.e. $r_{c}=\frac{\pi}{\sqrt{3}}\frac{\kb}{p_{loc}}$, which is rewritten as
\begin{equation}
\label{r_clocalized_rewritten}
r_{c}p_{F}/\kb=\frac{\pi}{\sqrt{3}}\frac{p_{F}}{p_{loc}}.
\end{equation}
Fig.~\ref{fig:pfrcvspfoverplock3435kbar2} shows a good agreement with Eq.~\eqref{r_clocalized_rewritten} and the extracted correlation length, estimated from the second moment of $g_1(r)$. The inset shows also that Eq.~\eqref{r_clocalized_rewritten} describes well the exponential decay of the coherence function for a given set of parameters.

\begin{figure}[t!]
\includegraphics[scale=0.2,clip]{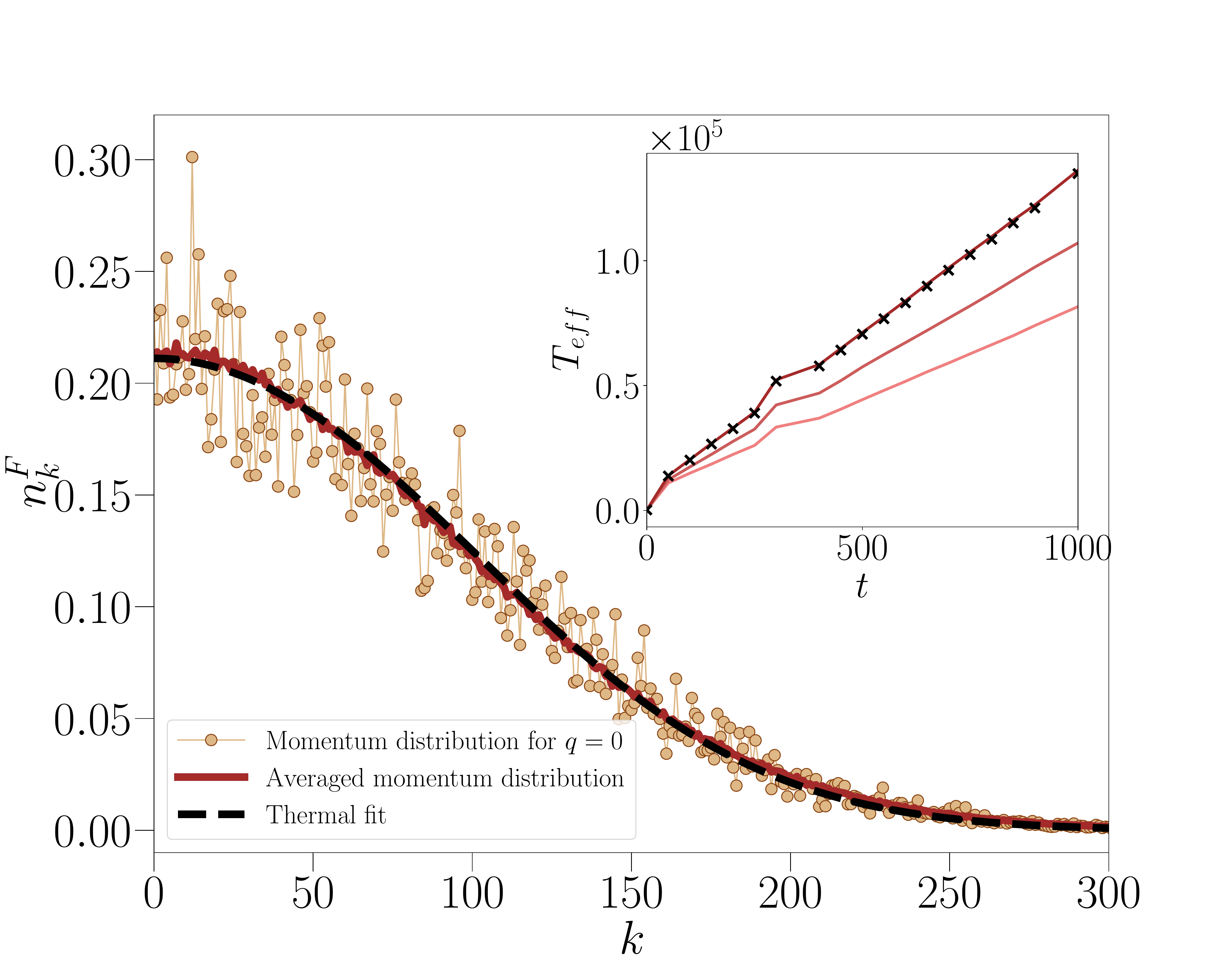} 
\caption{Averaged momentum distribution (continuous line) compared to the thermal fit with the Maxwell-Boltzmann distribution $f_{MB}$ (dashed line) deep in the delocalized phase. Here $N=51$, $K=12$, and $t=1000$ kicks and we averaged over 150 $q$. Inset: Time evolution of the effective temperature for $N=101$ and $K=12$, 14 and 16. The prediction $T_{eff}=\frac{2E}{N}$ is shown as black crosses for $K=16$.}
\label{fig:averageddistributiondeloc}
\end{figure}

\subsection{Delocalized phase\label{sub_sec_deloc}}

For large values of $(K,\varepsilon)$, the single-particle orbitals are delocalized, and characterized by a diffusive dynamics, and each of their energies grows linearly in time and the orbitals are Gaussian, see Fig.~\ref{fig:averageddistributiondeloc}. We keep $(\kb,\varepsilon)=(2.89,0.55)$ in the numerics.
Thus, the energy will scale as $E(t)\simeq E_{0} + 2N\mathcal{D}t$, as shown in Fig.~\ref{fig:localized energy} in red.  Note that at long enough times, each fermionic wave-function is much larger than the initial fermionic momentum distribution (of width $2p_F\simeq N$), and thus the momentum distribution of the $N$ fermions is just a Gaussian with increasing width. We focus on this asymptotic regime from now on.
We will show below that the particular shape of the momentum distribution of both fermions and bosons can be described by a high-temperature thermal gas with a temperature that increases linearly in time. The system is thus asymptotically reaching an infinite temperature phase.

\begin{figure}[t!]
\centering
\includegraphics[scale=0.2]{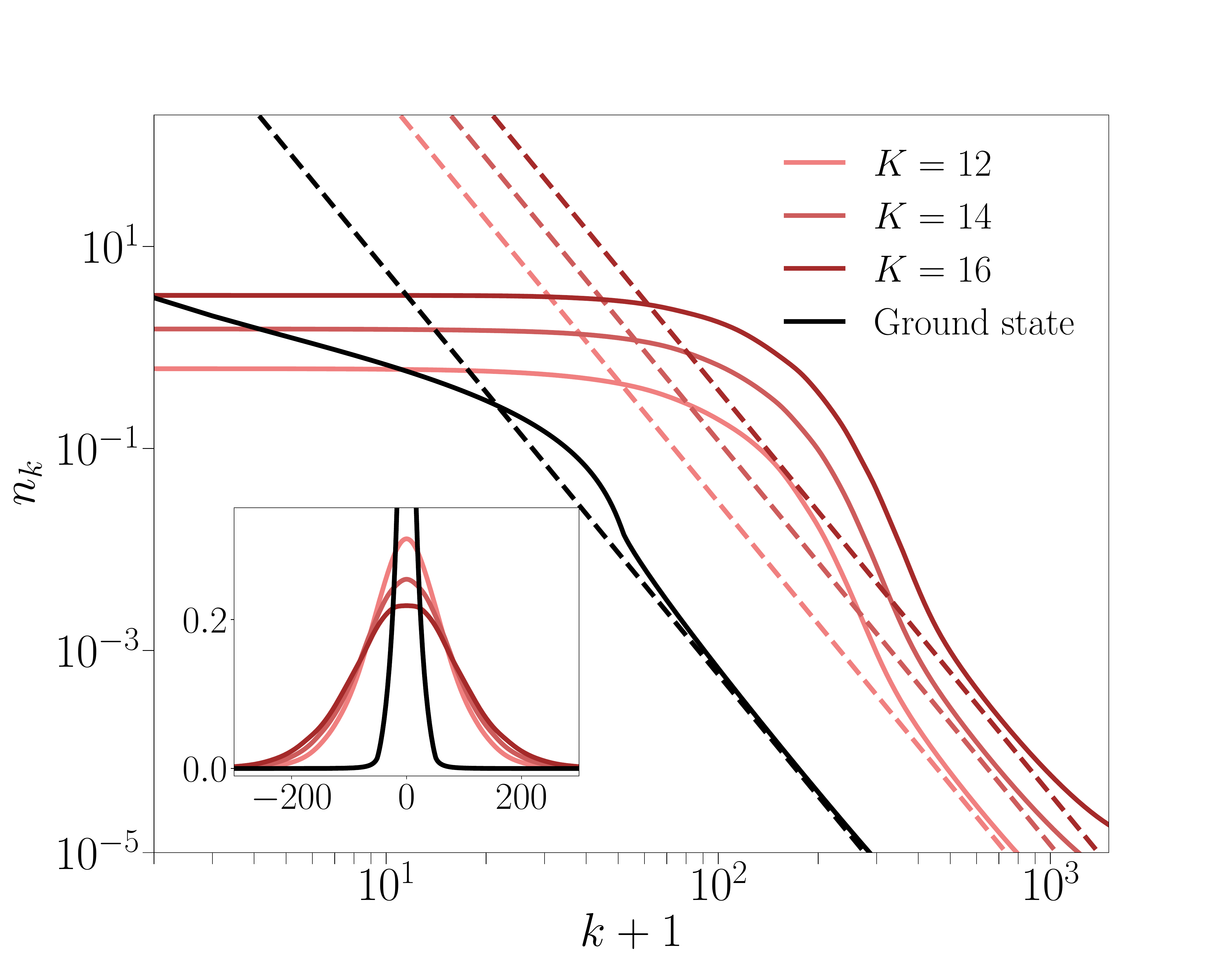}
\includegraphics[scale=0.2]{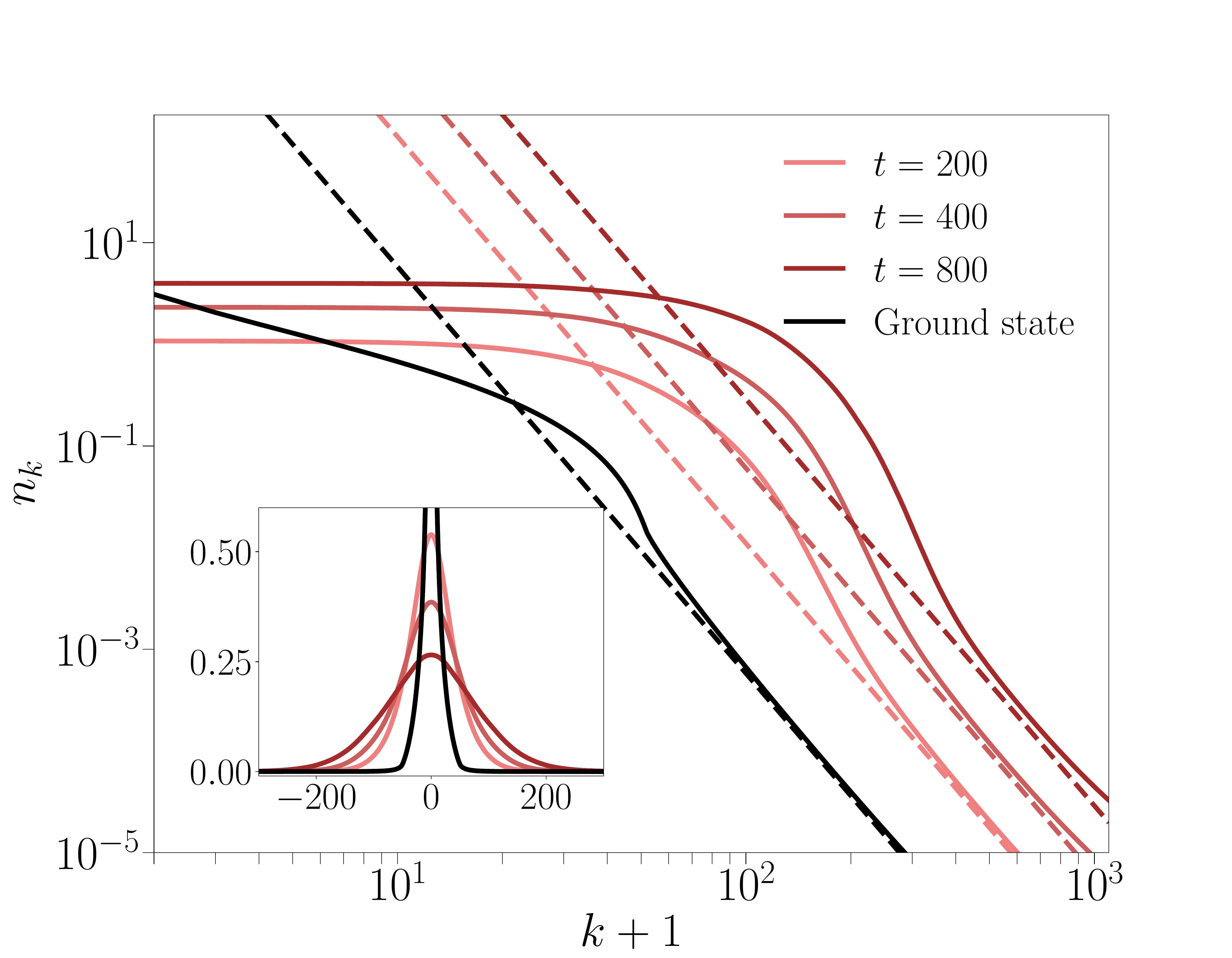}
\caption{
Momentum distribution deep in the delocalized phase for $N=51$ particles in log-log scale (for better visibility in both main panels, the different curves have been shifted  vertically by adding a constant to $n_k$, from top to bottom: 15, 6, 2, 0). Top panel: fixed time $t=600$ and various $K$. Bottom panel: fixed $K=12$ and various times. In both panels, the dashed line shows the asymptotic behavior $n_{k}\simeq \mathcal{C}_{th}/k^{4}$ at large momenta, with $\mathcal{C}_{th}$ computed from Eq.~\eqref{eqtanhighT}. The inset shows the same quantities in linear scale. }
\label{momentum distribution diffusive phase}
\end{figure}

\begin{figure}[h]
\centering
\includegraphics[scale=0.18,clip]{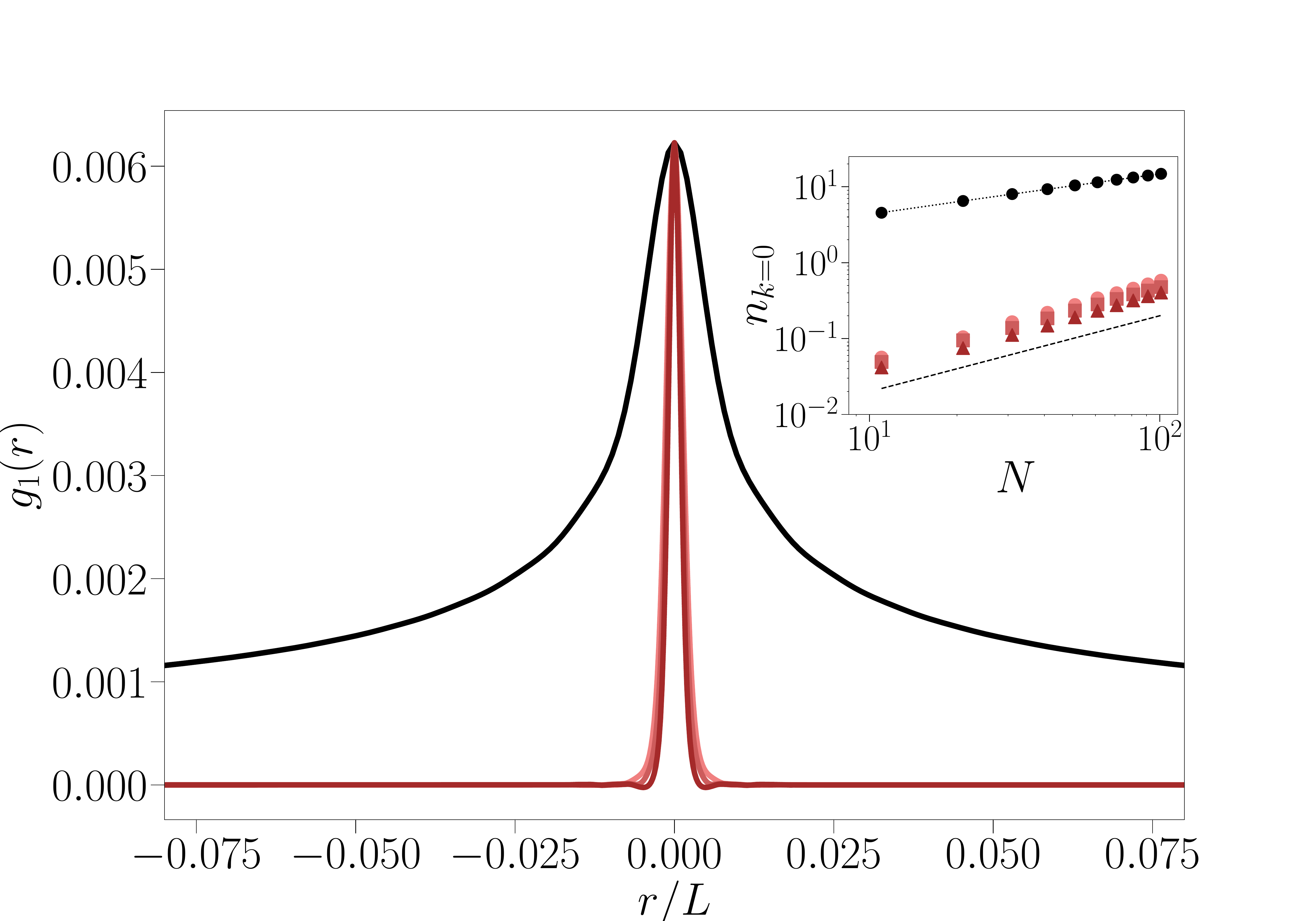}
\includegraphics[scale=0.18,clip]{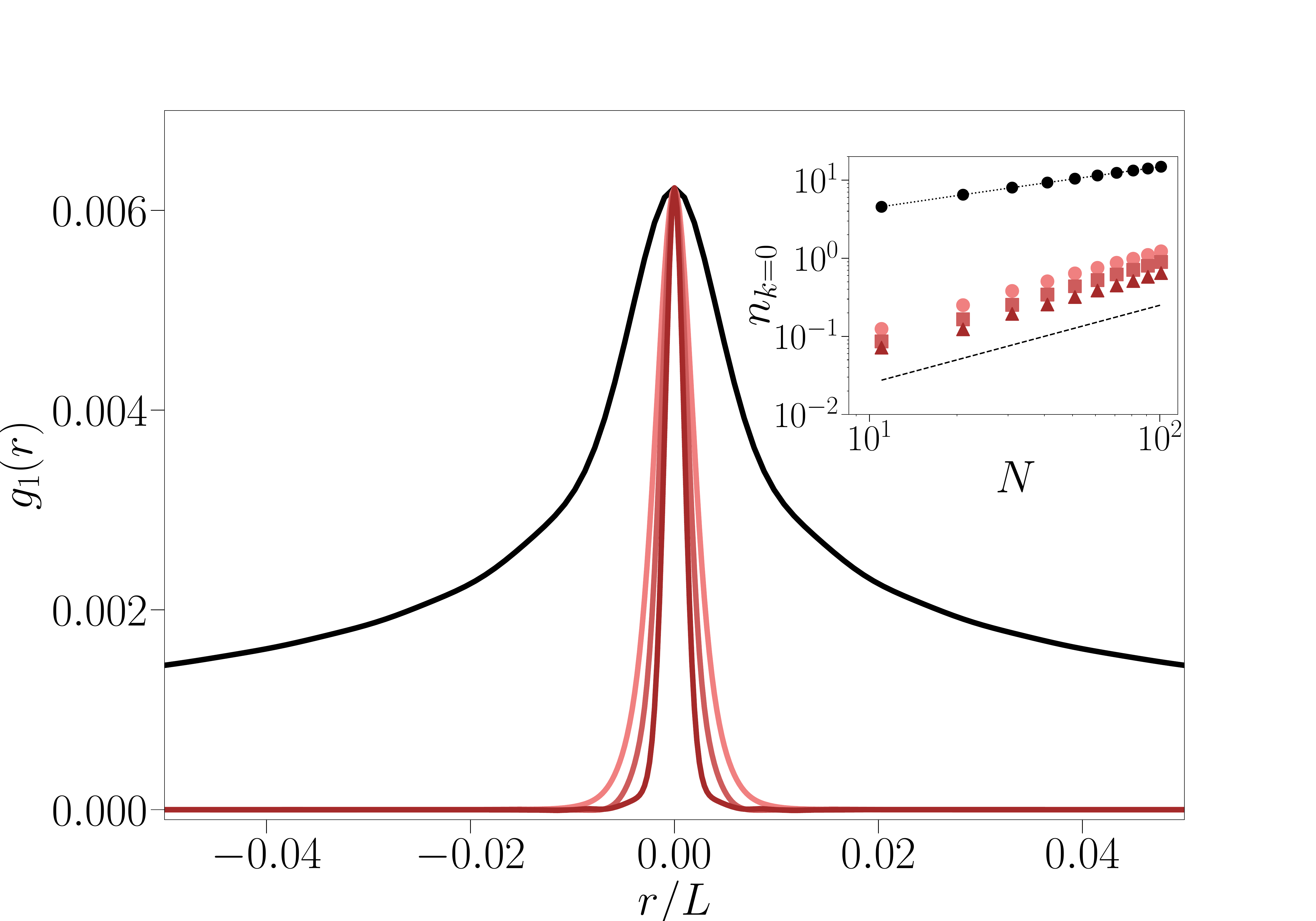}
\caption{Top panel: Coherence function $g_{1}(r)$ for $K=12$, 14 and 16, $t=600$. Bottom panel: Coherence function for $t=200$, 400 and 800, $K=12$. 
In both cases, $N=51$ and the legend is the same as in Fig.~\ref{momentum distribution diffusive phase}. For each panel, the inset shows the occupation of the zero-momentum state $n_{k=0}$. In both cases, $n_{k=0}$  scales linearly with $N$ in the delocalized phase (see dashed black line which is a guide to the eyes), while it grows as $\sqrt{N}$ in the ground state.}
\label{fig:density_matrix_deloc_phase_K_and_t}
\end{figure}

The momentum distribution of the Tonks gas at long time is shown in Fig.~\ref{momentum distribution diffusive phase}.  As in the localized phase, the divergence at small momenta present in the ground-state is destroyed by the kicks, while the asymptotic behavior at large momenta remains a power-law decay in $k^{-4}$. We plot again the coherence $g_{1}(r)$ in Fig.~\ref{fig:density_matrix_deloc_phase_K_and_t} as a function of $K$ (top panel) and $t$ (bottom panel). In both cases, a loss of coherence is seen since $g_{1}$ decay exponentially fast, eventually looking like a delta Dirac at very long time.  

We also stress that the occupation of $n_{k=0}$ scales linearly with the number of particles $N$, see right inset of Fig.~\ref{fig:density_matrix_deloc_phase_K_and_t}, which might be surprising at first sight, as it is usually interpreted (for instance in the ground state) as a hallmark of long-range coherence. However, as we show below, the system is in fact in a very high-temperature regime, and the bosons behave as a collection of $N$ almost independent particles (up to the contact physics at very high momenta). Hence the linear scaling in the number of particles of the momentum distribution (the same is observed for the fermions). 


As in the localized regime, we can extract an effective temperature and chemical potential for the fermions. In particular, we find that the effective chemical potential is large and negative, $|\mu_{eff}|/T_{eff}\gg 1$, and the Fermi-Dirac distribution is nothing but the Maxwell-Boltzmann distribution $f_{MB}$,
\begin{equation}
f_{MB}(k,T,\mu)\propto e^{-\frac{\kb^{2}k^{2}}{2T}+\mu/T},
\end{equation}
which is Gaussian, as is the momentum distribution of the fermions in the delocalized phase.  Using this distribution to fit fermionic distributions at long time works very well indeed, see Fig.~\ref{fig:averageddistributiondeloc}.
Note that since the energy grows with time, we expect the effective temperature and chemical potential to change in time too.
We can extract an effective temperature and an effective chemical potential as done in the localized regime, using the constraints 
\begin{equation}\label{eqfMB}\begin{split}
&\sum_{k\in\mathbb{Z}}f_{MB}(k,T_{eff},\mu_{eff})=N,\\
&\sum_{k\in\mathbb{Z}}\frac{\kb^{2}k^{2}}{2}f_{MB}(k,T_{eff},\mu_{eff})=E.
\end{split}
\end{equation}
The equipartition theorem implies that 
\begin{equation}
\label{teff_energy_n}
\begin{split}
T_{eff}&=\frac{2E}{N},\\
&=4\mathcal{D}t.
\end{split}
\end{equation}
We show in the inset of Fig.~\ref{fig:averageddistributiondeloc}, the time evolution of the extracted temperature for different $K$. It increases linearly with time, while the prediction $T_{eff}=2E/N$ works very well (black crosses).

\begin{figure}[t!]
\centering
\includegraphics[scale=0.20,clip]{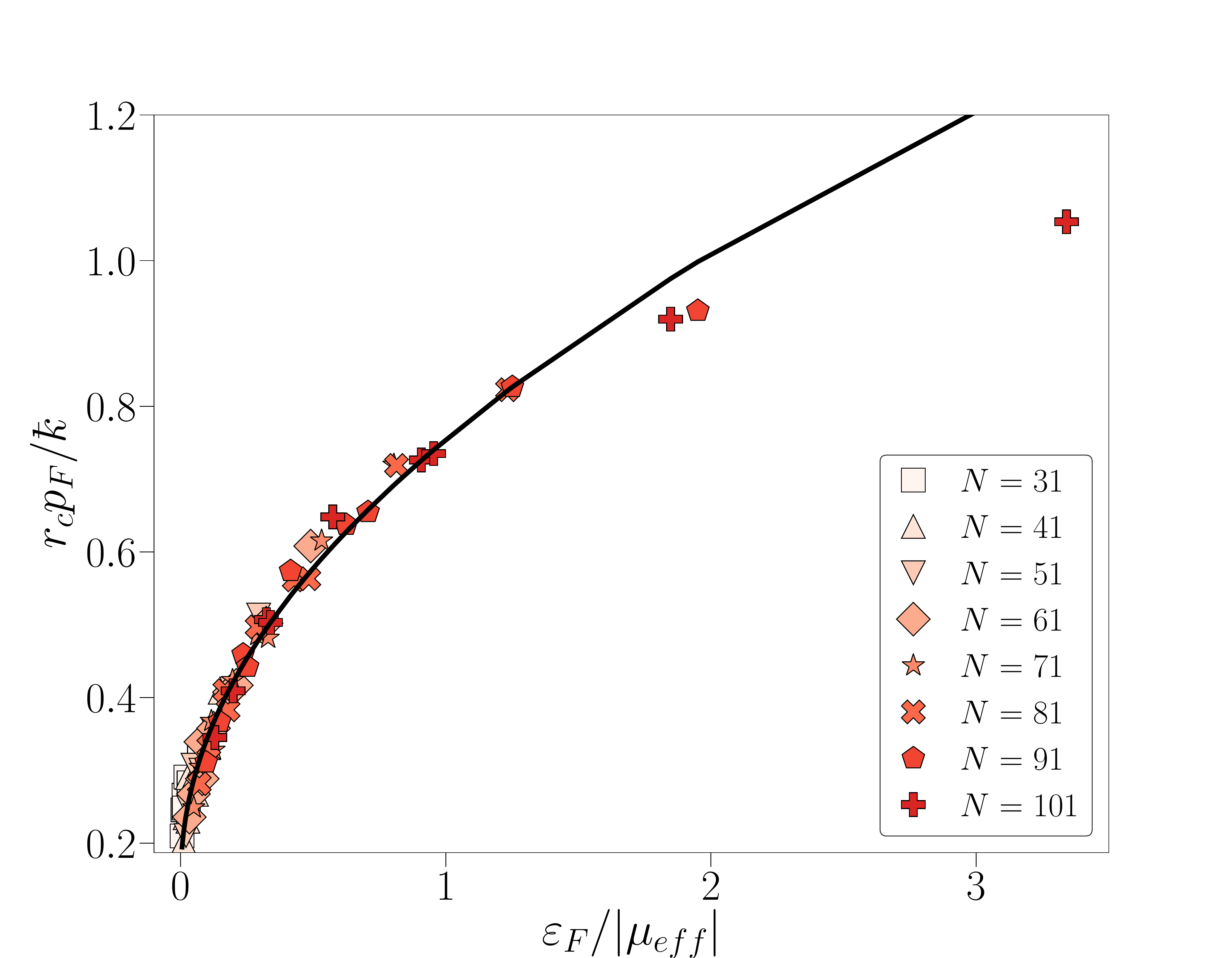} 
\caption{
Correlation length $r_{c}$ as a function of $\varepsilon_{F}/|\mu_{eff}|$ in the delocalized phase for various values of $N$, $t$ and $K$. The black line is the analytical prediction for a thermal Tonks gas at negative chemical potential, see text. } 
\label{fig:correlationlength}
\end{figure}

Using Eqs.~\eqref{eq_thermalcontact} and \eqref{teff_energy_n}, we obtain
\begin{equation}\label{eqtanhighT}
\mathcal{C}_{th}(N,T_{eff})=\frac{4N^{2}T_{eff}}{L^{2}\kb^{2}},
\end{equation}
which is consistent with what was found in the high-temperature limit \cite{Vignolo2013}, and indicates that the contact scales linearly with time. We show in Fig.~\ref{momentum distribution diffusive phase} as dashed lines that the power-law decay is indeed very well described by $\mathcal{C}_{th}(N,T_{eff})/k^{4}$.

Finally, the width of the coherence function, $r_c$, is given for a thermal Tonks gas with negative chemical potential by \cite{Its1991}
\begin{equation}\label{eq:debrogliewl}
r_{c}^{-1}=\frac{\sqrt{2|\mu|}}{\kb} R(\mu/T),
\end{equation}
with
\begin{equation}
\begin{split}
R(y) = 1+\frac{1}{2\pi\sqrt{|y|}}\int_{-\infty}^\infty dz \ln\left|\frac{e^{z^2-y}+1}{e^{z^2-y}-1}\right|.
\end{split}
\end{equation}
We, therefore, expect $r_c p_F/\kb$ to for various $N$ to be described by a universal function of $\varepsilon_F/|\mu|$. This is verified in Fig.~\ref{fig:correlationlength}. Equation \eqref{eq:debrogliewl}, which tends to $\frac{\sqrt{2|\mu|}}{\kb}$ at large $|\mu|/T$, describes very well the data for $|\mu|\gg \varepsilon_F$. 
 Nevertheless, we observe a departure from the analytical prediction at smaller chemical potentials --corresponding to very high effective temperatures--, which can be due to the estimation of $r_{c}$ by the second  moment of the coherence function, as it works only for intermediate temperatures \cite{Rigol2005temp}.

\subsection{Critical regime\label{sub_sec_crit}}

\begin{figure}[t!!!!!!]
    \centering
    \includegraphics[scale=0.2]{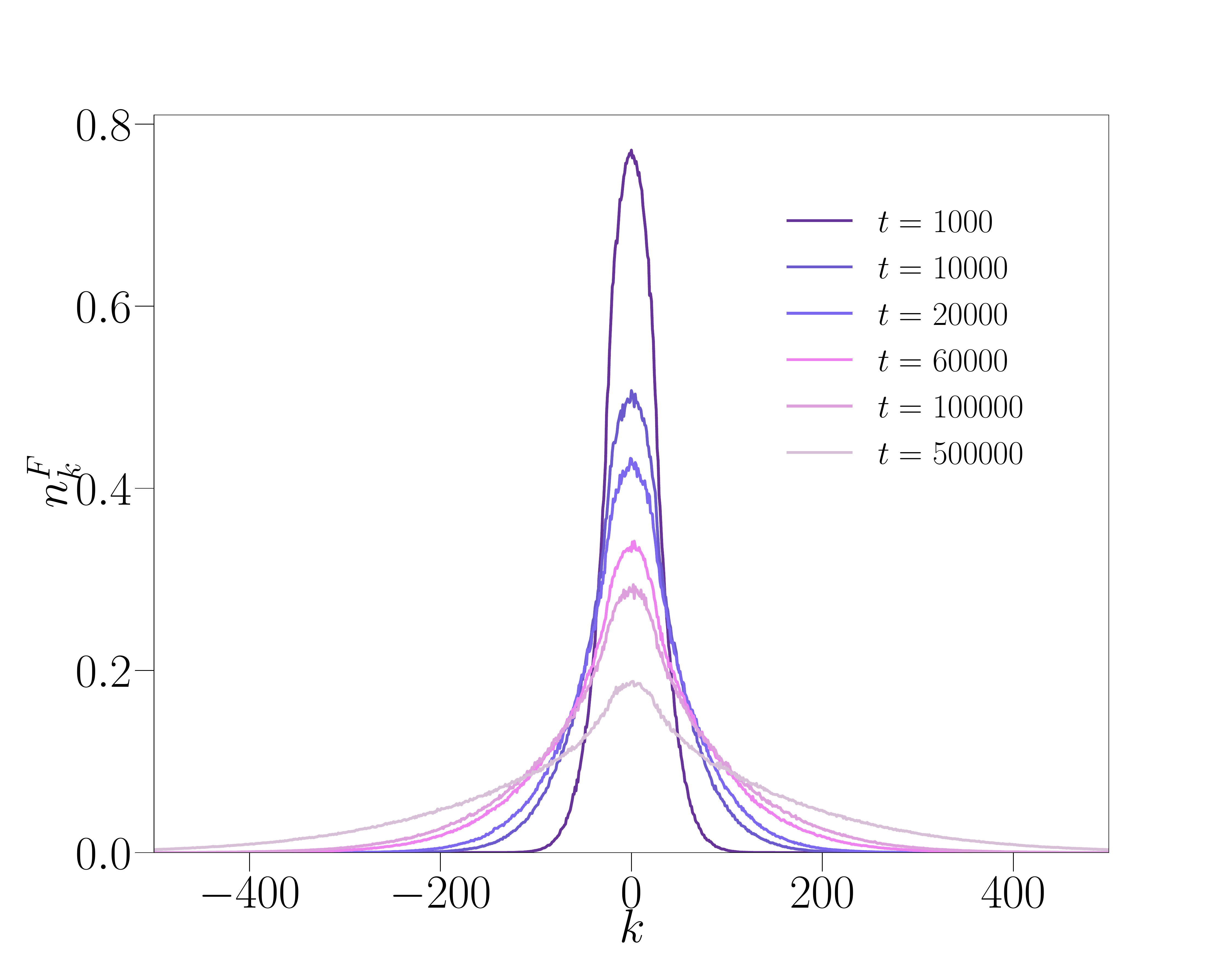}
    \includegraphics[scale=0.2,clip]{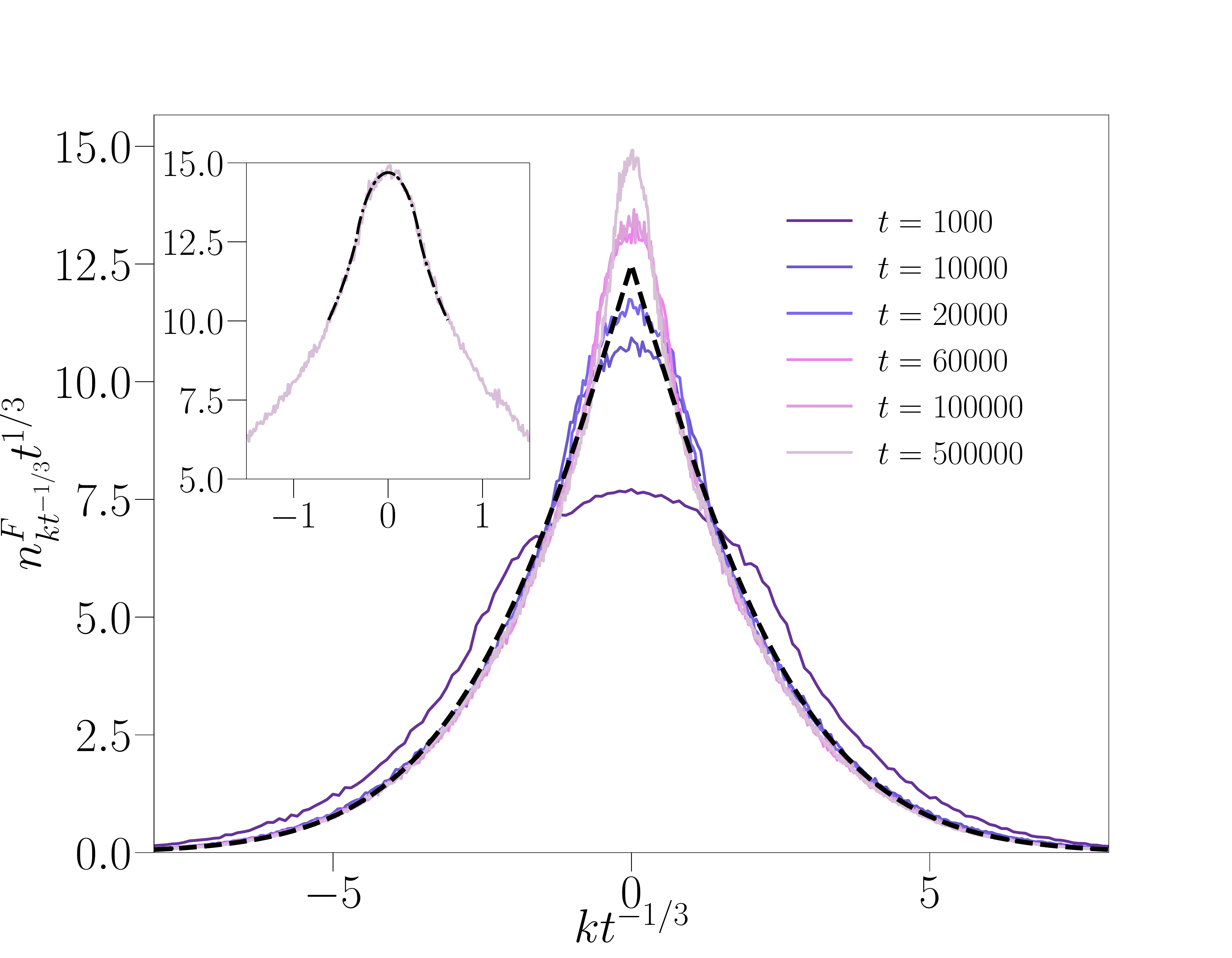}
    \caption{Top panel: Fermionic momentum distribution at criticality for various times, with $N=51$.
    Bottom panel: Same data once rescaled. The black dash line is the self-consistent theory prediction Eq.~\eqref{eq:selfconsistentmanybodyequation}. Inset: Inner part of the momentum distribution at $t=500000$, fitted by the multifractal singularity Eq.\eqref{eq:multifractal} convolved with the initial fermionic momentum distribution, see text. }
    \label{fig:fermionic many body critical distribution}
\end{figure}

To study the critical phase, we choose $(\kbar,K,\varepsilon)\simeq(2.85,6.36,0.43)$ \cite{Lemarie2009}, for which the energy of the system grows as $t^{2/3}$, see Fig.~\ref{fig:localized energy} in purple. We start by analyzing the free fermions. We have seen in Sec.~\ref{sec_model} that the single-particle orbitals are expected to obey the scaling form Eq.~\eqref{eq:scaling}, with a width that grows as $t^{1/3}$. Thus, at sufficiently long time, such that each orbital is much larger than the initial momentum distribution of the fermions, i.e. for $L_t\gg p_F$, we expect the fermionic momentum distribution to be given by the single-particle scaling function
\begin{equation}\label{eq:selfconsistentmanybodyequation}
\begin{split}
       n_{k}^{F}&=N  t^{-1/3}f(|k|t^{-1/3}),\\
    &\simeq N t^{-1/3}\frac{3^{2/3}}{2\gamma^{1/2}}{\rm Ai}\left(\frac{|kt^{-1/3}|}{3^{1/3}\gamma^{1/2}}\right),
\end{split}
\end{equation}
where on the second line we have used the result of the self-consistent theory, valid at large enough momenta.
This is verified in  Fig.~\ref{fig:fermionic many body critical distribution}, where we see that the rescaled fermionic momentum distribution is well described in the wings by Eq.~\eqref{eq:selfconsistentmanybodyequation} at sufficiently long times. Note that the central part of the distribution is not captured by the self-consistent theory, and grows in time to become the multifractal non-analyticity Eq.~\eqref{eq:multifractal}. The inset of the lower panel of Fig.~\ref{fig:fermionic many body critical distribution} shows that at long times, the central part of the momentum distribution is indeed well described by the multifractal peak, Eq.~\eqref{eq:multifractal} (convolved with the initial momentum distribution), with effective fractal dimension $d_{2,eff}=1.33 \pm 0.06$ at $t=5\times 10^{5}$ in agreement with Ref.~\cite{Panayotis2019}.  Therefore, for time large enough such that $L_t\gg p_F$, the momentum distribution of the fermions is described by the one-parameter scaling theory, as expected. This implies that the fermions cannot be described as a thermal gas, signaling a breaking of the effective thermalization observed in the localized and delocalized phases.

We are now in a position to  analyze the critical behavior of the Tonks gas. To minimize the numerical artifacts induced by the finite cut-off in momentum, we mostly plot data for $N=11$.
Naively, using the Bose-Fermi mapping, one could infer that the bosonic momentum distribution obeys the same scaling law as that of the fermions. However, as we will see, the picture is more complicated.

\begin{figure}[t!!!!!!!!]
    \centering
    \includegraphics[scale=0.2,clip]{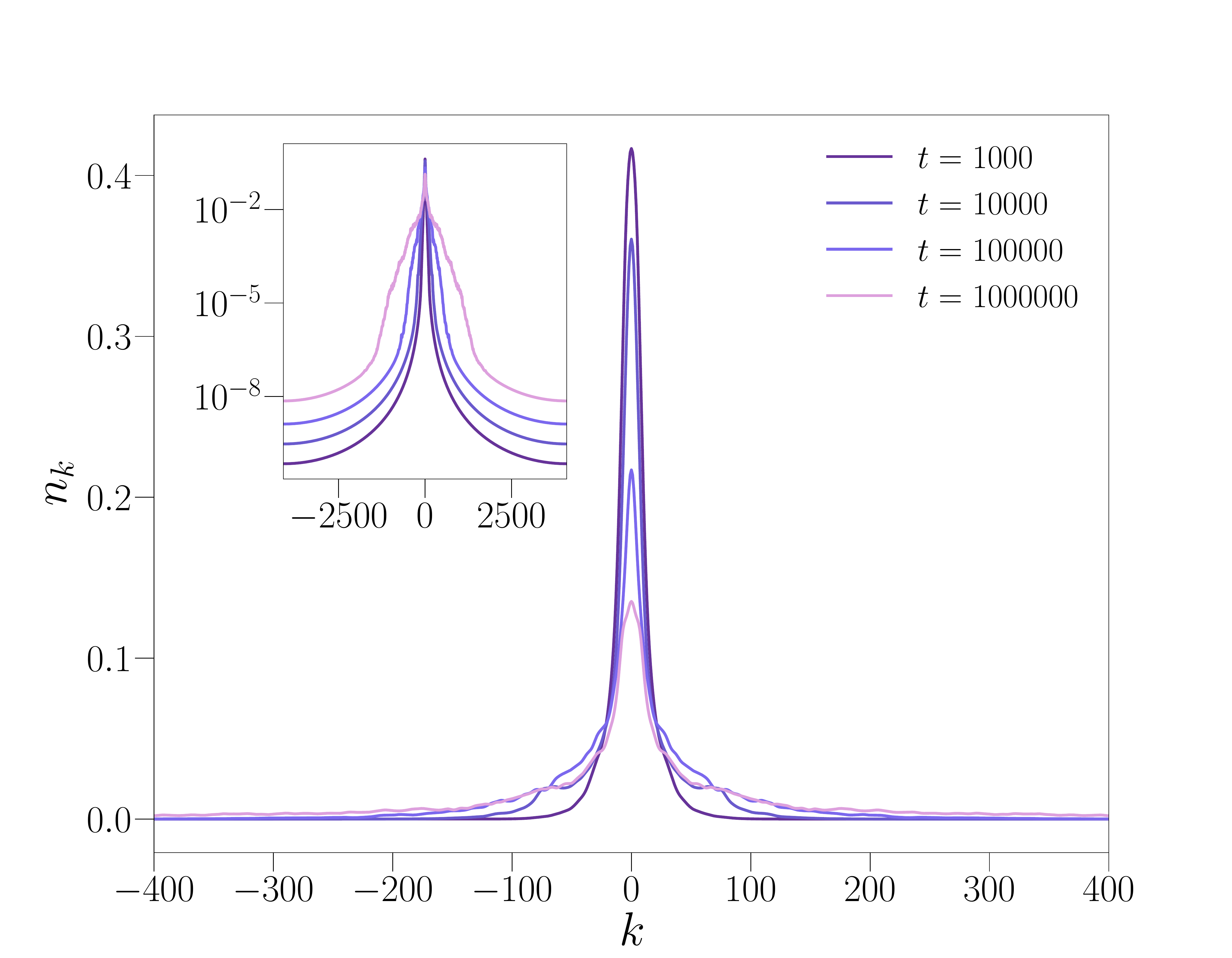}
    \includegraphics[scale=0.2,clip]{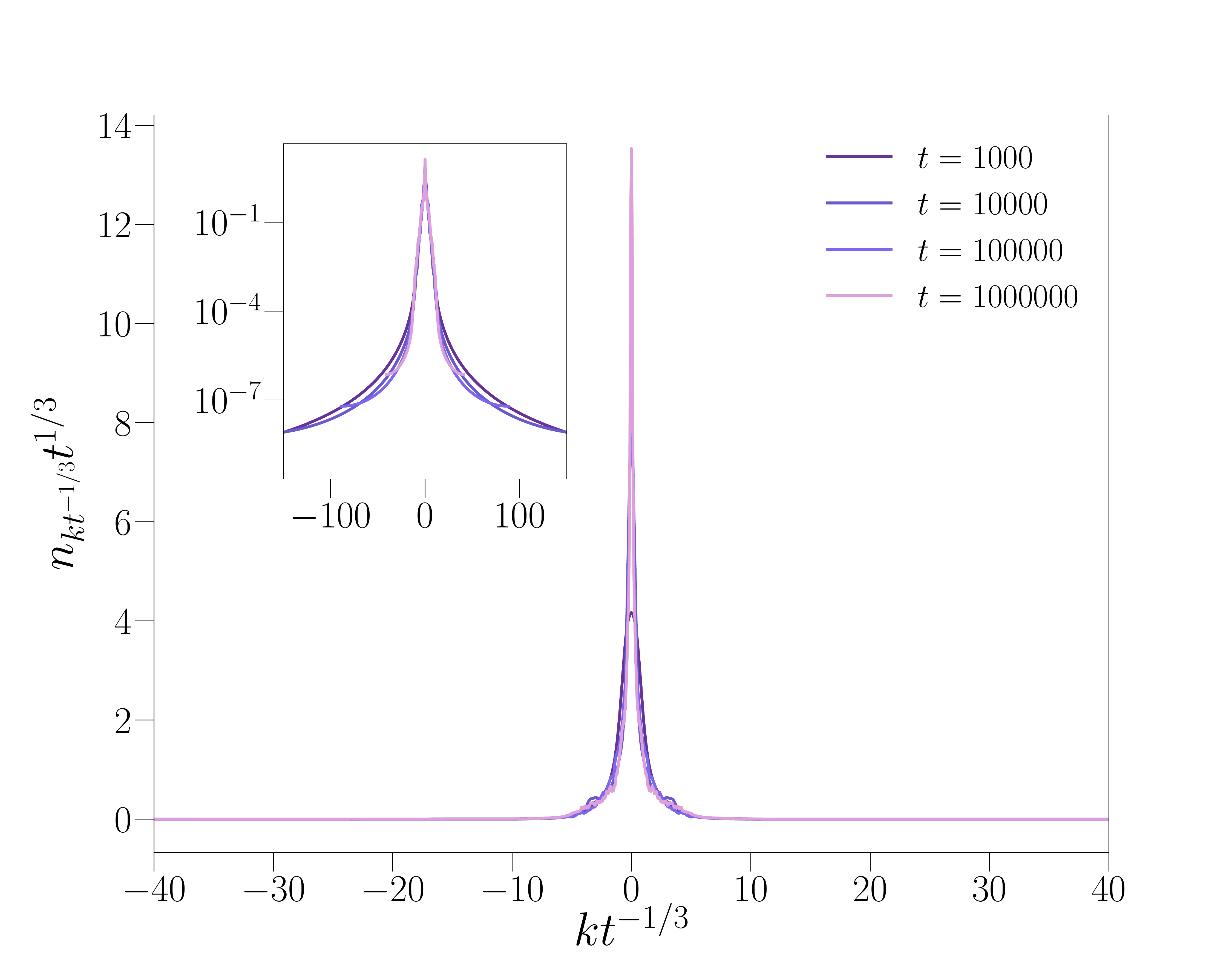}
    \caption{Top panel: bosonic momentum distribution at criticality for different times for $N=11$.
    Bottom panel: same data once rescaled. Both insets show the data of the corresponding main panel in semi-log scale.}
    \label{fig:bosonic momentum distribution critic regime}
\end{figure}

\begin{figure}[t!!!!]
    \centering
    \includegraphics[scale=0.2,clip]{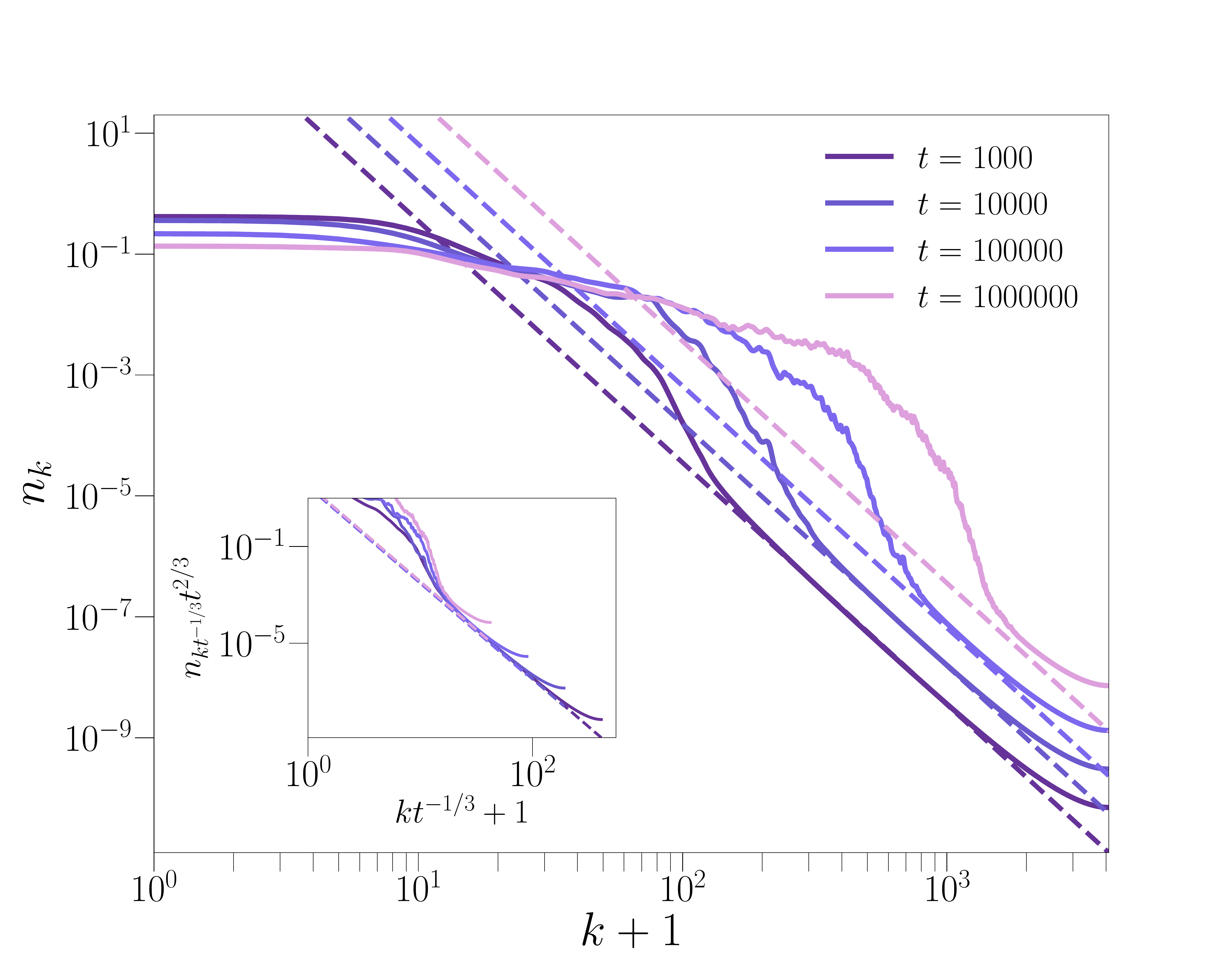}
    \caption{Main panel: Same data as Fig.~\ref{fig:bosonic momentum distribution critic regime} but in log-log scale. Note that the momentum distributions have \textit{not} been shifted. Dashed lines represent the relation $\mathcal{C}=8NE/L^{2}\kb^{2}$, which explains very well the distribution's behavior at large momenta. Inset: same data as in the main panel, once rescaled (see main text).}
    \label{fig:loglog_critical_momentum_distribution_bosons}
\end{figure}

Fig.~\ref{fig:bosonic momentum distribution critic regime} shows that for moderate momenta (the behavior close to zero-momentum will be discussed below), the bosonic momentum distribution obeys the standard scaling law,
\begin{equation}
\label{eq_scaling1}
    n_k(t)=t^{-1/3}f_1(k t^{-1/3}).
\end{equation}
We note however that the data do not obey the scaling law at large enough momenta. Fig.~\ref{fig:loglog_critical_momentum_distribution_bosons} focuses on that region in momentum space. As expected for an interacting quantum system, and as we have already seen in the localized and delocalized regimes, the momentum distribution decays as $1/k^4$, with a contact that increases in time. In fact, using once more the formula $\mathcal{C}=8NE/L^{2}\kb^{2}$, we find that the contact scales as $t^{2/3}$. Therefore, we find that for large momenta, the momentum distribution obeys a scaling law of the form
\begin{equation}
\label{eq_scaling2}
    n_k(t)=t^{-2/3}f_2(k t^{-1/3}),
\end{equation}
with $f_2(y)=y^{-4}$, see inset of Fig.~\ref{fig:loglog_critical_momentum_distribution_bosons}. This inset shows in particular that the cross-over scale $k_c(t)$ between the two regimes scales as $t^{1/3}$. 

\begin{figure}[t!!!!!]
    \centering
    \includegraphics[scale=0.2,clip]{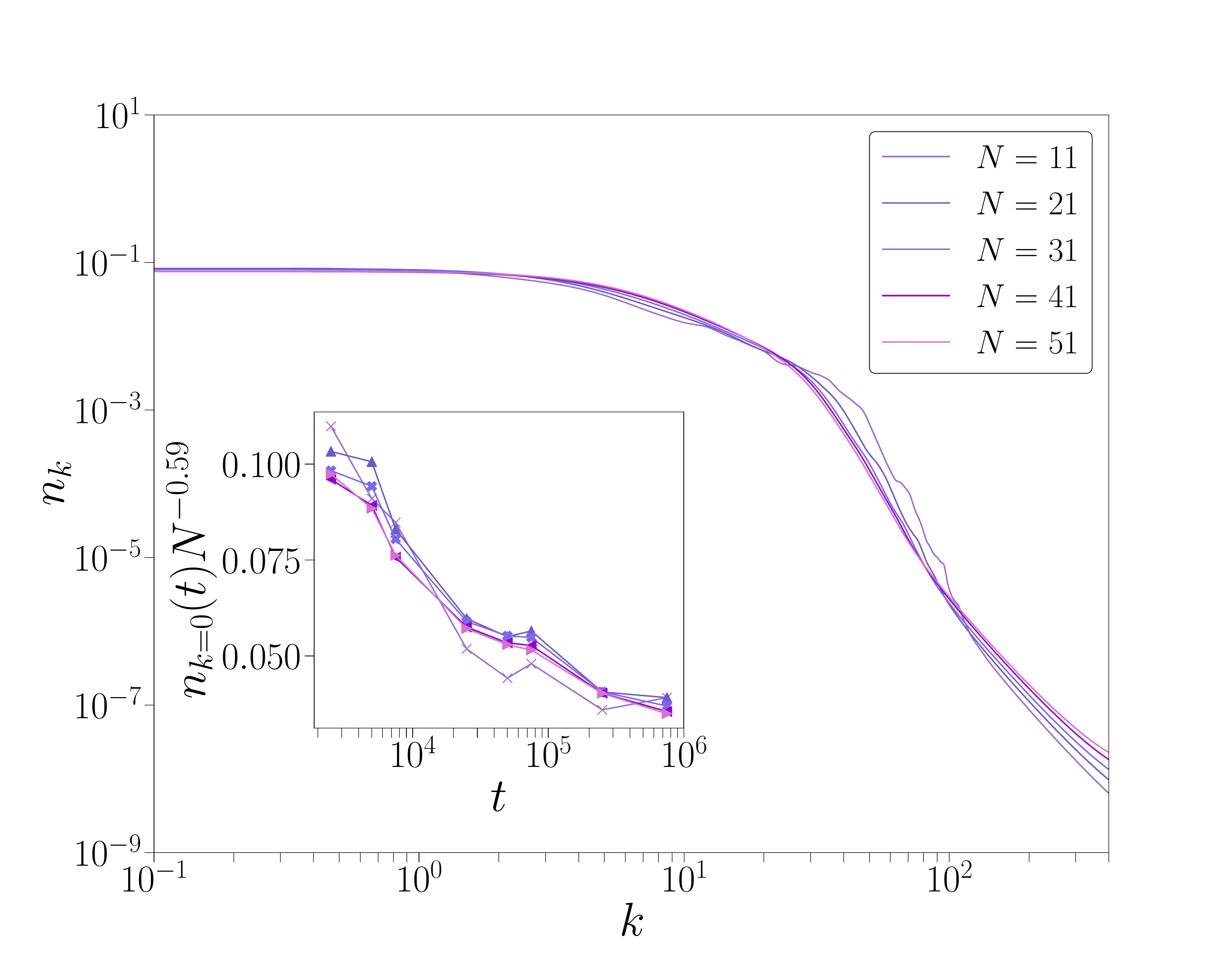}
    \caption{Main panel: same data as Fig.~\ref{fig:loglog_critical_momentum_distribution_bosons} but rescaled with the number of particles (see text) for various $N$. We observe a collapse of the data for momenta smaller than a typical momentum of order $N^{1/3}$. Inset: time evolution of  central peak $n_{k=0}$ rescaled by $N^{-0.59}$.
    }
    \label{fig:main_panel_n_k_resc_N059_vs_k_resc_N033_inset_n_k=0_resc_N059_vs_t}
\end{figure}

On the other hand, the inset of Fig.~\ref{fig:main_panel_n_k_resc_N059_vs_k_resc_N033_inset_n_k=0_resc_N059_vs_t}  shows that $n_{k=0}(t)$ scales with the number of particles as $N^\gamma$, with exponent $\gamma\simeq 0.59$. As for the delocalized phase, this should not be interpreted in favor of (quasi) long-range coherence, although we do not have yet an explanation for this behavior. Then, assuming that for small enough momenta $n_k(t)$ also scales as $N^{0.59}$, we are able to find a collapse of the data for various $N$ at fixed time by rescaling the momenta by $N^{1/3}$, see Fig.~\ref{fig:main_panel_n_k_resc_N059_vs_k_resc_N033_inset_n_k=0_resc_N059_vs_t} main panel. We thus infer that the momentum scale $k_c(t)$ between the scaling forms Eq.~\eqref{eq_scaling1} and Eq.~\eqref{eq_scaling2} is of order $(Nt)^{1/3}\propto (p_F t)^{1/3}$ (to be contrasted with the ground state, for which the cross-over scale between small and large momentum behavior is given by $p_F\sim N$). This implies a clear breaking of the one-parameter scaling law  expected for the Anderson transition of non-interacting particles. A similar breaking has been observed in the quasi-periodically kicked Gross-Pitaevski equation in \cite{Cherroret2014}.

Finally, we have analyzed the behavior of the central peak at small momenta of the bosonic momentum distribution, and we have not found any sign of multifractality, as the bosonic momentum distribution appears to be regular (quadratic) at small momenta. Whether multifractality can be observed in the quasi-periodically kicked Tonks gas is left for future work.

\section{Discussion\label{conclusion}}
We have studied the different phases of the interacting QPQKR in the Tonks limit. Both the localized and delocalized phases have been shown to behave  effectively like a thermal gas, respectively in the low-temperature and high-temperature regimes. In both cases, we have shown that the bosonic distribution is drastically different from the fermionic one, and decay at large momenta as a power law given by a Tan's contact, which depends on the temperature in accordance with what was found in Ref.~\cite{Vignolo2013}. Even though it provides a quantitative description of both of these phases, thermalization remains effective, which implies that the natural orbitals of the one-body density matrix are not plane waves, see App.~\ref{Natorb}. Finally, we made a quantitative analysis of the critical regime at the transition. 
While the phase diagram is the same as that of non-interacting particles, the momentum distribution of the Tonks gas does not display the one-parameter scaling law. Indeed, there is a cross-over between a small momentum regime scaling as  $t^{1/3}$ (similar to free particles) and a large momentum regime scaling as $t^{2/3}$ (dominated by the scaling of the contact). The cross-over scale between these two regimes appears to scale as $(p_F t)^{1/3}$. Contrary to the non-interacting case, we have not been able to identify the effects of multifractality in the behavior of the momentum distribution, which might nevertheless influence other observables.

Beyond the Tonks regime, it would be interesting to investigate a quasi-periodic kicked Lieb-Liniger model to see whether the transition is preserved by finite interactions, and in particular if the critical exponents are modified. Previous studies in the mean-field (Gross-Pitaevskii) regime suggest that a transition is still present, and characterized by a new critical exponent, while the localized phase is replaced by subdiffusive transport \cite{Cherroret2014}. It is a very interesting question to understand how beyond mean-field effects would change this picture.


\section*{Acknowledgments}
We thank R. Chicireanu, J.-C. Garreau, D. Delande, N. Cherroret for  discussions. This work was supported by Agence Nationale de la Recherche through Research Grants QRITiC I-SITE ULNE/ ANR-16-IDEX-0004 ULNE, the Labex CEMPI Grant No.ANR-11-LABX-0007-01, the Programme Investissements d'Avenir ANR-11-IDEX-0002-02, reference ANR-10-LABX-0037-NEXT and the Ministry of Higher Education and Research, Hauts-de-France Council and European Regional Development Fund (ERDF) through the Contrat de Projets \'Etat-Region (CPER Photonics for Society, P4S).

\appendix
\section{Natural orbitals and finite-time scaling\label{Natorb}}

\begin{figure}[h!!!!!!!!!!]
\centering
\includegraphics[scale=0.195,clip]{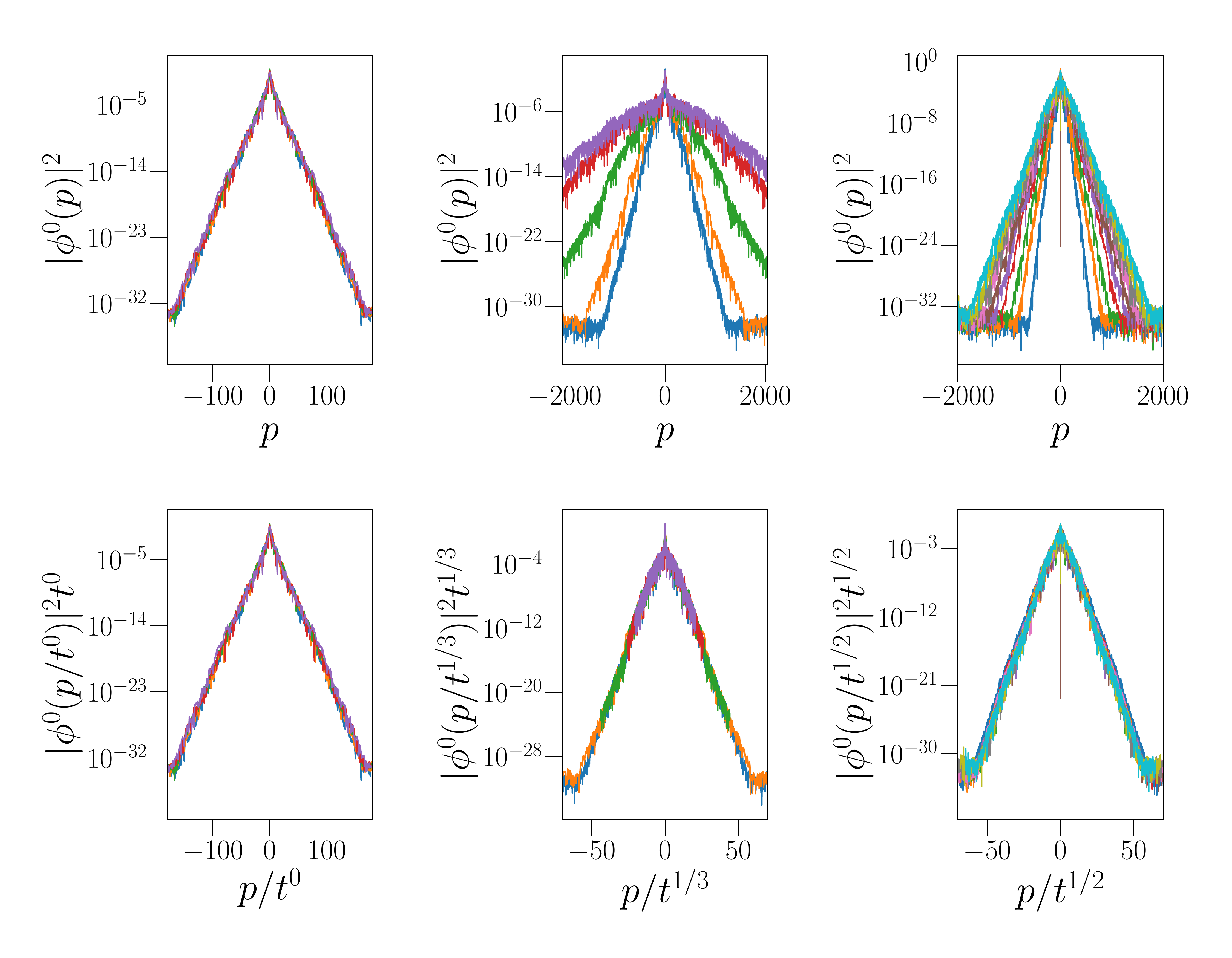}
\caption{
Most occupied natural orbital in the three regimes from left to right: localized, critical, and delocalized. The lower panels show the same data as the upper panel, but using rescaled variables that highlight the scaling behavior. Parameters used here are : $t=1100$, 1200, 1300, 1400, 1500 in the localized regime ($\kb=2.89$, $K=3.5$,  $\varepsilon=0.55$); $t=$ 10000, 20000, 100000, 500000, 1000000 in the critical regime ($\kb=2.85$, $K=6.36$, $\varepsilon=0.43$); and  $t\in [100,1000]$ in steps of 100 in diffusive regime ($\kb=2.89$, $K=12$, $\varepsilon=0.55$).}
\label{fig:mono_finite_time_scaling}
\end{figure}

The one-body density matrix can be decomposed in natural orbitals, noted $\phi^{\eta}(x)$, which informs on the modes occupied by the particles in the context of many-body physics. They are defined as
\begin{equation}
\int dy\rho(x,y)\phi^{\eta}(y)=\lambda_{\eta}\phi^{\eta}(x),
\end{equation}
with $\lambda_{\eta}$ the occupation number of the $\eta$-th natural orbital. We focus here on the dynamics of the most occupied natural orbital  in momentum space $\phi^{0}(p)$ in the three regimes (localized, diffusive, critical), see Fig.~\ref{fig:mono_finite_time_scaling}. The upper panel shows the dynamics, with clearly distinct behaviors in the three regimes. The lower panel uses the expected rescaling associated with each regime (in particular with $t^{1/3}$ for the critical regime, as in the non-interacting case) and shows a very good collapse at all times.

\bibliography{bibliography.bib,revtex-custom.bib}

\begin{thebibliography}{56}%
\makeatletter
\providecommand \@ifxundefined [1]{%
 \@ifx{#1\undefined}
}%
\providecommand \@ifnum [1]{%
 \ifnum #1\expandafter \@firstoftwo
 \else \expandafter \@secondoftwo
 \fi
}%
\providecommand \@ifx [1]{%
 \ifx #1\expandafter \@firstoftwo
 \else \expandafter \@secondoftwo
 \fi
}%
\providecommand \natexlab [1]{#1}%
\providecommand \enquote  [1]{``#1''}%
\providecommand \bibnamefont  [1]{#1}%
\providecommand \bibfnamefont [1]{#1}%
\providecommand \citenamefont [1]{#1}%
\providecommand \href@noop [0]{\@secondoftwo}%
\providecommand \href [0]{\begingroup \@sanitize@url \@href}%
\providecommand \@href[1]{\@@startlink{#1}\@@href}%
\providecommand \@@href[1]{\endgroup#1\@@endlink}%
\providecommand \@sanitize@url [0]{\catcode `\\12\catcode `\$12\catcode
  `\&12\catcode `\#12\catcode `\^12\catcode `\_12\catcode `\%12\relax}%
\providecommand \@@startlink[1]{}%
\providecommand \@@endlink[0]{}%
\providecommand \url  [0]{\begingroup\@sanitize@url \@url }%
\providecommand \@url [1]{\endgroup\@href {#1}{\urlprefix }}%
\providecommand \urlprefix  [0]{URL }%
\providecommand \Eprint [0]{\href }%
\providecommand \doibase [0]{http://dx.doi.org/}%
\providecommand \selectlanguage [0]{\@gobble}%
\providecommand \bibinfo  [0]{\@secondoftwo}%
\providecommand \bibfield  [0]{\@secondoftwo}%
\providecommand \translation [1]{[#1]}%
\providecommand \BibitemOpen [0]{}%
\providecommand \bibitemStop [0]{}%
\providecommand \bibitemNoStop [0]{.\EOS\space}%
\providecommand \EOS [0]{\spacefactor3000\relax}%
\providecommand \BibitemShut  [1]{\csname bibitem#1\endcsname}%
\let\auto@bib@innerbib\@empty
\bibitem [{\citenamefont {Anderson}(1958)}]{Anderson1958}%
  \BibitemOpen
  \bibfield  {author} {\bibinfo {author} {\bibfnamefont {P.~W.}\ \bibnamefont
  {Anderson}},\ }\href {\doibase 10.1103/PhysRev.109.1492} {\bibfield
  {journal} {\bibinfo  {journal} {Phys. Rev.}\ }\textbf {\bibinfo {volume}
  {109}},\ \bibinfo {pages} {1492} (\bibinfo {year} {1958})}\BibitemShut
  {NoStop}%
\bibitem [{\citenamefont {Evers}\ and\ \citenamefont
  {Mirlin}(2008)}]{Evers2008}%
  \BibitemOpen
  \bibfield  {author} {\bibinfo {author} {\bibfnamefont {F.}~\bibnamefont
  {Evers}}\ and\ \bibinfo {author} {\bibfnamefont {A.~D.}\ \bibnamefont
  {Mirlin}},\ }\href {\doibase 10.1103/RevModPhys.80.1355} {\bibfield
  {journal} {\bibinfo  {journal} {Rev. Mod. Phys.}\ }\textbf {\bibinfo {volume}
  {80}},\ \bibinfo {pages} {1355} (\bibinfo {year} {2008})}\BibitemShut
  {NoStop}%
\bibitem [{\citenamefont {Fishman}\ \emph {et~al.}(1982)\citenamefont
  {Fishman}, \citenamefont {Grempel},\ and\ \citenamefont
  {Prange}}]{Fishman1982}%
  \BibitemOpen
  \bibfield  {author} {\bibinfo {author} {\bibfnamefont {S.}~\bibnamefont
  {Fishman}}, \bibinfo {author} {\bibfnamefont {D.~R.}\ \bibnamefont
  {Grempel}}, \ and\ \bibinfo {author} {\bibfnamefont {R.~E.}\ \bibnamefont
  {Prange}},\ }\href {\doibase 10.1103/PhysRevLett.49.509} {\bibfield
  {journal} {\bibinfo  {journal} {Phys. Rev. Lett.}\ }\textbf {\bibinfo
  {volume} {49}},\ \bibinfo {pages} {509} (\bibinfo {year} {1982})}\BibitemShut
  {NoStop}%
\bibitem [{\citenamefont {Shepelyansky}(1987)}]{Shepelyansky1987}%
  \BibitemOpen
  \bibfield  {author} {\bibinfo {author} {\bibfnamefont {D.}~\bibnamefont
  {Shepelyansky}},\ }\href {\doibase
  https://doi.org/10.1016/0167-2789(87)90123-0} {\bibfield  {journal} {\bibinfo
   {journal} {Physica D: Nonlinear Phenomena}\ }\textbf {\bibinfo {volume}
  {28}},\ \bibinfo {pages} {103} (\bibinfo {year} {1987})}\BibitemShut
  {NoStop}%
\bibitem [{\citenamefont {Casati}\ \emph {et~al.}(1989)\citenamefont {Casati},
  \citenamefont {Guarneri},\ and\ \citenamefont {Shepelyansky}}]{Casati1989}%
  \BibitemOpen
  \bibfield  {author} {\bibinfo {author} {\bibfnamefont {G.}~\bibnamefont
  {Casati}}, \bibinfo {author} {\bibfnamefont {I.}~\bibnamefont {Guarneri}}, \
  and\ \bibinfo {author} {\bibfnamefont {D.~L.}\ \bibnamefont {Shepelyansky}},\
  }\href {\doibase 10.1103/PhysRevLett.62.345} {\bibfield  {journal} {\bibinfo
  {journal} {Phys. Rev. Lett.}\ }\textbf {\bibinfo {volume} {62}},\ \bibinfo
  {pages} {345} (\bibinfo {year} {1989})}\BibitemShut {NoStop}%
\bibitem [{\citenamefont {Chab\'e}\ \emph {et~al.}(2008)\citenamefont
  {Chab\'e}, \citenamefont {Lemari\'e}, \citenamefont {Gr\'emaud},
  \citenamefont {Delande}, \citenamefont {Szriftgiser},\ and\ \citenamefont
  {Garreau}}]{Chabe2008}%
  \BibitemOpen
  \bibfield  {author} {\bibinfo {author} {\bibfnamefont {J.}~\bibnamefont
  {Chab\'e}}, \bibinfo {author} {\bibfnamefont {G.}~\bibnamefont {Lemari\'e}},
  \bibinfo {author} {\bibfnamefont {B.}~\bibnamefont {Gr\'emaud}}, \bibinfo
  {author} {\bibfnamefont {D.}~\bibnamefont {Delande}}, \bibinfo {author}
  {\bibfnamefont {P.}~\bibnamefont {Szriftgiser}}, \ and\ \bibinfo {author}
  {\bibfnamefont {J.~C.}\ \bibnamefont {Garreau}},\ }\href {\doibase
  10.1103/PhysRevLett.101.255702} {\bibfield  {journal} {\bibinfo  {journal}
  {Phys. Rev. Lett.}\ }\textbf {\bibinfo {volume} {101}},\ \bibinfo {pages}
  {255702} (\bibinfo {year} {2008})}\BibitemShut {NoStop}%
\bibitem [{\citenamefont {Lemari\'e}\ \emph {et~al.}(2009)\citenamefont
  {Lemari\'e}, \citenamefont {Chab\'e}, \citenamefont {Szriftgiser},
  \citenamefont {Garreau}, \citenamefont {Gr\'emaud},\ and\ \citenamefont
  {Delande}}]{Lemarie2009}%
  \BibitemOpen
  \bibfield  {author} {\bibinfo {author} {\bibfnamefont {G.}~\bibnamefont
  {Lemari\'e}}, \bibinfo {author} {\bibfnamefont {J.}~\bibnamefont {Chab\'e}},
  \bibinfo {author} {\bibfnamefont {P.}~\bibnamefont {Szriftgiser}}, \bibinfo
  {author} {\bibfnamefont {J.~C.}\ \bibnamefont {Garreau}}, \bibinfo {author}
  {\bibfnamefont {B.}~\bibnamefont {Gr\'emaud}}, \ and\ \bibinfo {author}
  {\bibfnamefont {D.}~\bibnamefont {Delande}},\ }\href {\doibase
  10.1103/PhysRevA.80.043626} {\bibfield  {journal} {\bibinfo  {journal} {Phys.
  Rev. A}\ }\textbf {\bibinfo {volume} {80}},\ \bibinfo {pages} {043626}
  (\bibinfo {year} {2009})}\BibitemShut {NoStop}%
\bibitem [{\citenamefont {Lemari\'e}\ \emph {et~al.}(2010)\citenamefont
  {Lemari\'e}, \citenamefont {Lignier}, \citenamefont {Delande}, \citenamefont
  {Szriftgiser},\ and\ \citenamefont {Garreau}}]{Lemarie2010}%
  \BibitemOpen
  \bibfield  {author} {\bibinfo {author} {\bibfnamefont {G.}~\bibnamefont
  {Lemari\'e}}, \bibinfo {author} {\bibfnamefont {H.}~\bibnamefont {Lignier}},
  \bibinfo {author} {\bibfnamefont {D.}~\bibnamefont {Delande}}, \bibinfo
  {author} {\bibfnamefont {P.}~\bibnamefont {Szriftgiser}}, \ and\ \bibinfo
  {author} {\bibfnamefont {J.~C.}\ \bibnamefont {Garreau}},\ }\href {\doibase
  10.1103/PhysRevLett.105.090601} {\bibfield  {journal} {\bibinfo  {journal}
  {Phys. Rev. Lett.}\ }\textbf {\bibinfo {volume} {105}},\ \bibinfo {pages}
  {090601} (\bibinfo {year} {2010})}\BibitemShut {NoStop}%
\bibitem [{\citenamefont {Lopez}\ \emph {et~al.}(2012)\citenamefont {Lopez},
  \citenamefont {Cl\'ement}, \citenamefont {Szriftgiser}, \citenamefont
  {Garreau},\ and\ \citenamefont {Delande}}]{Lopez2012}%
  \BibitemOpen
  \bibfield  {author} {\bibinfo {author} {\bibfnamefont {M.}~\bibnamefont
  {Lopez}}, \bibinfo {author} {\bibfnamefont {J.-F.}\ \bibnamefont
  {Cl\'ement}}, \bibinfo {author} {\bibfnamefont {P.}~\bibnamefont
  {Szriftgiser}}, \bibinfo {author} {\bibfnamefont {J.~C.}\ \bibnamefont
  {Garreau}}, \ and\ \bibinfo {author} {\bibfnamefont {D.}~\bibnamefont
  {Delande}},\ }\href {\doibase 10.1103/PhysRevLett.108.095701} {\bibfield
  {journal} {\bibinfo  {journal} {Phys. Rev. Lett.}\ }\textbf {\bibinfo
  {volume} {108}},\ \bibinfo {pages} {095701} (\bibinfo {year}
  {2012})}\BibitemShut {NoStop}%
\bibitem [{\citenamefont {Manai}\ \emph {et~al.}(2015)\citenamefont {Manai},
  \citenamefont {Cl\'ement}, \citenamefont {Chicireanu}, \citenamefont
  {Hainaut}, \citenamefont {Garreau}, \citenamefont {Szriftgiser},\ and\
  \citenamefont {Delande}}]{Manai2015}%
  \BibitemOpen
  \bibfield  {author} {\bibinfo {author} {\bibfnamefont {I.}~\bibnamefont
  {Manai}}, \bibinfo {author} {\bibfnamefont {J.-F.}\ \bibnamefont
  {Cl\'ement}}, \bibinfo {author} {\bibfnamefont {R.}~\bibnamefont
  {Chicireanu}}, \bibinfo {author} {\bibfnamefont {C.}~\bibnamefont {Hainaut}},
  \bibinfo {author} {\bibfnamefont {J.~C.}\ \bibnamefont {Garreau}}, \bibinfo
  {author} {\bibfnamefont {P.}~\bibnamefont {Szriftgiser}}, \ and\ \bibinfo
  {author} {\bibfnamefont {D.}~\bibnamefont {Delande}},\ }\href {\doibase
  10.1103/PhysRevLett.115.240603} {\bibfield  {journal} {\bibinfo  {journal}
  {Phys. Rev. Lett.}\ }\textbf {\bibinfo {volume} {115}},\ \bibinfo {pages}
  {240603} (\bibinfo {year} {2015})}\BibitemShut {NoStop}%
\bibitem [{\citenamefont {Nandkishore}(2015)}]{Nandkishore2015}%
  \BibitemOpen
  \bibfield  {author} {\bibinfo {author} {\bibfnamefont {R.}~\bibnamefont
  {Nandkishore}},\ }\href {\doibase 10.1103/PhysRevB.92.245141} {\bibfield
  {journal} {\bibinfo  {journal} {Phys. Rev. B}\ }\textbf {\bibinfo {volume}
  {92}},\ \bibinfo {pages} {245141} (\bibinfo {year} {2015})}\BibitemShut
  {NoStop}%
\bibitem [{\citenamefont {Abanin}\ \emph {et~al.}(2019)\citenamefont {Abanin},
  \citenamefont {Altman}, \citenamefont {Bloch},\ and\ \citenamefont
  {Serbyn}}]{Abanin2019}%
  \BibitemOpen
  \bibfield  {author} {\bibinfo {author} {\bibfnamefont {D.~A.}\ \bibnamefont
  {Abanin}}, \bibinfo {author} {\bibfnamefont {E.}~\bibnamefont {Altman}},
  \bibinfo {author} {\bibfnamefont {I.}~\bibnamefont {Bloch}}, \ and\ \bibinfo
  {author} {\bibfnamefont {M.}~\bibnamefont {Serbyn}},\ }\href {\doibase
  10.1103/RevModPhys.91.021001} {\bibfield  {journal} {\bibinfo  {journal}
  {Rev. Mod. Phys.}\ }\textbf {\bibinfo {volume} {91}},\ \bibinfo {pages}
  {021001} (\bibinfo {year} {2019})}\BibitemShut {NoStop}%
\bibitem [{\citenamefont {Serbyn}\ \emph {et~al.}(2013)\citenamefont {Serbyn},
  \citenamefont {Papi\ifmmode~\acute{c}\else \'{c}\fi{}},\ and\ \citenamefont
  {Abanin}}]{Serbyn2013}%
  \BibitemOpen
  \bibfield  {author} {\bibinfo {author} {\bibfnamefont {M.}~\bibnamefont
  {Serbyn}}, \bibinfo {author} {\bibfnamefont {Z.}~\bibnamefont
  {Papi\ifmmode~\acute{c}\else \'{c}\fi{}}}, \ and\ \bibinfo {author}
  {\bibfnamefont {D.~A.}\ \bibnamefont {Abanin}},\ }\href {\doibase
  10.1103/PhysRevLett.111.127201} {\bibfield  {journal} {\bibinfo  {journal}
  {Phys. Rev. Lett.}\ }\textbf {\bibinfo {volume} {111}},\ \bibinfo {pages}
  {127201} (\bibinfo {year} {2013})}\BibitemShut {NoStop}%
\bibitem [{\citenamefont {Huse}\ \emph {et~al.}(2014)\citenamefont {Huse},
  \citenamefont {Nandkishore},\ and\ \citenamefont {Oganesyan}}]{Huse2014}%
  \BibitemOpen
  \bibfield  {author} {\bibinfo {author} {\bibfnamefont {D.~A.}\ \bibnamefont
  {Huse}}, \bibinfo {author} {\bibfnamefont {R.}~\bibnamefont {Nandkishore}}, \
  and\ \bibinfo {author} {\bibfnamefont {V.}~\bibnamefont {Oganesyan}},\ }\href
  {\doibase 10.1103/PhysRevB.90.174202} {\bibfield  {journal} {\bibinfo
  {journal} {Phys. Rev. B}\ }\textbf {\bibinfo {volume} {90}},\ \bibinfo
  {pages} {174202} (\bibinfo {year} {2014})}\BibitemShut {NoStop}%
\bibitem [{\citenamefont {Ponte}\ \emph
  {et~al.}(2015{\natexlab{a}})\citenamefont {Ponte}, \citenamefont
  {Papi\ifmmode~\acute{c}\else \'{c}\fi{}}, \citenamefont {Huveneers},\ and\
  \citenamefont {Abanin}}]{Ponte2015a}%
  \BibitemOpen
  \bibfield  {author} {\bibinfo {author} {\bibfnamefont {P.}~\bibnamefont
  {Ponte}}, \bibinfo {author} {\bibfnamefont {Z.}~\bibnamefont
  {Papi\ifmmode~\acute{c}\else \'{c}\fi{}}}, \bibinfo {author} {\bibfnamefont
  {F.~m.~c.}\ \bibnamefont {Huveneers}}, \ and\ \bibinfo {author}
  {\bibfnamefont {D.~A.}\ \bibnamefont {Abanin}},\ }\href {\doibase
  10.1103/PhysRevLett.114.140401} {\bibfield  {journal} {\bibinfo  {journal}
  {Phys. Rev. Lett.}\ }\textbf {\bibinfo {volume} {114}},\ \bibinfo {pages}
  {140401} (\bibinfo {year} {2015}{\natexlab{a}})}\BibitemShut {NoStop}%
\bibitem [{\citenamefont {Ponte}\ \emph
  {et~al.}(2015{\natexlab{b}})\citenamefont {Ponte}, \citenamefont {Chandran},
  \citenamefont {Papić},\ and\ \citenamefont {Abanin}}]{Ponte2015}%
  \BibitemOpen
  \bibfield  {author} {\bibinfo {author} {\bibfnamefont {P.}~\bibnamefont
  {Ponte}}, \bibinfo {author} {\bibfnamefont {A.}~\bibnamefont {Chandran}},
  \bibinfo {author} {\bibfnamefont {Z.}~\bibnamefont {Papić}}, \ and\ \bibinfo
  {author} {\bibfnamefont {D.~A.}\ \bibnamefont {Abanin}},\ }\href {\doibase
  https://doi.org/10.1016/j.aop.2014.11.008} {\bibfield  {journal} {\bibinfo
  {journal} {Annals of Physics}\ }\textbf {\bibinfo {volume} {353}},\ \bibinfo
  {pages} {196} (\bibinfo {year} {2015}{\natexlab{b}})}\BibitemShut {NoStop}%
\bibitem [{\citenamefont {Adachi}\ \emph {et~al.}(1988)\citenamefont {Adachi},
  \citenamefont {Toda},\ and\ \citenamefont {Ikeda}}]{Adachi1988}%
  \BibitemOpen
  \bibfield  {author} {\bibinfo {author} {\bibfnamefont {S.}~\bibnamefont
  {Adachi}}, \bibinfo {author} {\bibfnamefont {M.}~\bibnamefont {Toda}}, \ and\
  \bibinfo {author} {\bibfnamefont {K.}~\bibnamefont {Ikeda}},\ }\href
  {\doibase 10.1103/PhysRevLett.61.659} {\bibfield  {journal} {\bibinfo
  {journal} {Phys. Rev. Lett.}\ }\textbf {\bibinfo {volume} {61}},\ \bibinfo
  {pages} {659} (\bibinfo {year} {1988})}\BibitemShut {NoStop}%
\bibitem [{\citenamefont {Wen-Lei}\ and\ \citenamefont
  {Quan-Lin}(2009)}]{Lei2009}%
  \BibitemOpen
  \bibfield  {author} {\bibinfo {author} {\bibfnamefont {Z.}~\bibnamefont
  {Wen-Lei}}\ and\ \bibinfo {author} {\bibfnamefont {J.}~\bibnamefont
  {Quan-Lin}},\ }\href {\doibase 10.1088/0253-6102/51/3/17} {\bibfield
  {journal} {\bibinfo  {journal} {Communications in Theoretical Physics}\
  }\textbf {\bibinfo {volume} {51}},\ \bibinfo {pages} {465} (\bibinfo {year}
  {2009})}\BibitemShut {NoStop}%
\bibitem [{\citenamefont {Keser}\ \emph {et~al.}(2016)\citenamefont {Keser},
  \citenamefont {Ganeshan}, \citenamefont {Refael},\ and\ \citenamefont
  {Galitski}}]{Keser2016}%
  \BibitemOpen
  \bibfield  {author} {\bibinfo {author} {\bibfnamefont {A.~C.}\ \bibnamefont
  {Keser}}, \bibinfo {author} {\bibfnamefont {S.}~\bibnamefont {Ganeshan}},
  \bibinfo {author} {\bibfnamefont {G.}~\bibnamefont {Refael}}, \ and\ \bibinfo
  {author} {\bibfnamefont {V.}~\bibnamefont {Galitski}},\ }\href {\doibase
  10.1103/PhysRevB.94.085120} {\bibfield  {journal} {\bibinfo  {journal} {Phys.
  Rev. B}\ }\textbf {\bibinfo {volume} {94}},\ \bibinfo {pages} {085120}
  (\bibinfo {year} {2016})}\BibitemShut {NoStop}%
\bibitem [{\citenamefont {Rozenbaum}\ and\ \citenamefont
  {Galitski}(2017)}]{Rozenbaum2017}%
  \BibitemOpen
  \bibfield  {author} {\bibinfo {author} {\bibfnamefont {E.~B.}\ \bibnamefont
  {Rozenbaum}}\ and\ \bibinfo {author} {\bibfnamefont {V.}~\bibnamefont
  {Galitski}},\ }\href {\doibase 10.1103/PhysRevB.95.064303} {\bibfield
  {journal} {\bibinfo  {journal} {Phys. Rev. B}\ }\textbf {\bibinfo {volume}
  {95}},\ \bibinfo {pages} {064303} (\bibinfo {year} {2017})}\BibitemShut
  {NoStop}%
\bibitem [{\citenamefont {Notarnicola}\ \emph {et~al.}(2018)\citenamefont
  {Notarnicola}, \citenamefont {Iemini}, \citenamefont {Rossini}, \citenamefont
  {Fazio}, \citenamefont {Silva},\ and\ \citenamefont
  {Russomanno}}]{Notarnicola2018}%
  \BibitemOpen
  \bibfield  {author} {\bibinfo {author} {\bibfnamefont {S.}~\bibnamefont
  {Notarnicola}}, \bibinfo {author} {\bibfnamefont {F.}~\bibnamefont {Iemini}},
  \bibinfo {author} {\bibfnamefont {D.}~\bibnamefont {Rossini}}, \bibinfo
  {author} {\bibfnamefont {R.}~\bibnamefont {Fazio}}, \bibinfo {author}
  {\bibfnamefont {A.}~\bibnamefont {Silva}}, \ and\ \bibinfo {author}
  {\bibfnamefont {A.}~\bibnamefont {Russomanno}},\ }\href {\doibase
  10.1103/PhysRevE.97.022202} {\bibfield  {journal} {\bibinfo  {journal} {Phys.
  Rev. E}\ }\textbf {\bibinfo {volume} {97}},\ \bibinfo {pages} {022202}
  (\bibinfo {year} {2018})}\BibitemShut {NoStop}%
\bibitem [{\citenamefont {Notarnicola}\ \emph {et~al.}(2020)\citenamefont
  {Notarnicola}, \citenamefont {Silva}, \citenamefont {Fazio},\ and\
  \citenamefont {Russomanno}}]{Notarnicola2020}%
  \BibitemOpen
  \bibfield  {author} {\bibinfo {author} {\bibfnamefont {S.}~\bibnamefont
  {Notarnicola}}, \bibinfo {author} {\bibfnamefont {A.}~\bibnamefont {Silva}},
  \bibinfo {author} {\bibfnamefont {R.}~\bibnamefont {Fazio}}, \ and\ \bibinfo
  {author} {\bibfnamefont {A.}~\bibnamefont {Russomanno}},\ }\href {\doibase
  10.1088/1742-5468/ab6de4} {\bibfield  {journal} {\bibinfo  {journal} {Journal
  of Statistical Mechanics: Theory and Experiment}\ }\textbf {\bibinfo {volume}
  {2020}},\ \bibinfo {pages} {024008} (\bibinfo {year} {2020})}\BibitemShut
  {NoStop}%
\bibitem [{\citenamefont {Qin}\ \emph {et~al.}(2017)\citenamefont {Qin},
  \citenamefont {Andreanov}, \citenamefont {Park},\ and\ \citenamefont
  {Flach}}]{Qin2017}%
  \BibitemOpen
  \bibfield  {author} {\bibinfo {author} {\bibfnamefont {P.}~\bibnamefont
  {Qin}}, \bibinfo {author} {\bibfnamefont {A.}~\bibnamefont {Andreanov}},
  \bibinfo {author} {\bibfnamefont {H.~C.}\ \bibnamefont {Park}}, \ and\
  \bibinfo {author} {\bibfnamefont {S.}~\bibnamefont {Flach}},\ }\href
  {\doibase 10.1038/srep41139} {\bibfield  {journal} {\bibinfo  {journal}
  {Scientific Reports}\ }\textbf {\bibinfo {volume} {7}},\ \bibinfo {pages}
  {41139} (\bibinfo {year} {2017})}\BibitemShut {NoStop}%
\bibitem [{\citenamefont {Rylands}\ \emph {et~al.}(2020)\citenamefont
  {Rylands}, \citenamefont {Rozenbaum}, \citenamefont {Galitski},\ and\
  \citenamefont {Konik}}]{Rylands2020}%
  \BibitemOpen
  \bibfield  {author} {\bibinfo {author} {\bibfnamefont {C.}~\bibnamefont
  {Rylands}}, \bibinfo {author} {\bibfnamefont {E.~B.}\ \bibnamefont
  {Rozenbaum}}, \bibinfo {author} {\bibfnamefont {V.}~\bibnamefont {Galitski}},
  \ and\ \bibinfo {author} {\bibfnamefont {R.}~\bibnamefont {Konik}},\ }\href
  {\doibase 10.1103/PhysRevLett.124.155302} {\bibfield  {journal} {\bibinfo
  {journal} {Phys. Rev. Lett.}\ }\textbf {\bibinfo {volume} {124}},\ \bibinfo
  {pages} {155302} (\bibinfo {year} {2020})}\BibitemShut {NoStop}%
\bibitem [{\citenamefont {Chicireanu}\ and\ \citenamefont
  {Ran\ifmmode~\mbox{\c{c}}\else \c{c}\fi{}on}(2021)}]{Chicireanu2021}%
  \BibitemOpen
  \bibfield  {author} {\bibinfo {author} {\bibfnamefont {R.}~\bibnamefont
  {Chicireanu}}\ and\ \bibinfo {author} {\bibfnamefont {A.}~\bibnamefont
  {Ran\ifmmode~\mbox{\c{c}}\else \c{c}\fi{}on}},\ }\href {\doibase
  10.1103/PhysRevA.103.043314} {\bibfield  {journal} {\bibinfo  {journal}
  {Phys. Rev. A}\ }\textbf {\bibinfo {volume} {103}},\ \bibinfo {pages}
  {043314} (\bibinfo {year} {2021})}\BibitemShut {NoStop}%
\bibitem [{\citenamefont {Vuatelet}\ and\ \citenamefont
  {Ran\ifmmode~\mbox{\c{c}}\else \c{c}\fi{}on}(2021)}]{Vuatelet2021}%
  \BibitemOpen
  \bibfield  {author} {\bibinfo {author} {\bibfnamefont {V.}~\bibnamefont
  {Vuatelet}}\ and\ \bibinfo {author} {\bibfnamefont {A.}~\bibnamefont
  {Ran\ifmmode~\mbox{\c{c}}\else \c{c}\fi{}on}},\ }\href {\doibase
  10.1103/PhysRevA.104.043302} {\bibfield  {journal} {\bibinfo  {journal}
  {Phys. Rev. A}\ }\textbf {\bibinfo {volume} {104}},\ \bibinfo {pages}
  {043302} (\bibinfo {year} {2021})}\BibitemShut {NoStop}%
\bibitem [{\citenamefont {Shepelyansky}(1993)}]{Shepelyansky1993}%
  \BibitemOpen
  \bibfield  {author} {\bibinfo {author} {\bibfnamefont {D.~L.}\ \bibnamefont
  {Shepelyansky}},\ }\href {\doibase 10.1103/PhysRevLett.70.1787} {\bibfield
  {journal} {\bibinfo  {journal} {Phys. Rev. Lett.}\ }\textbf {\bibinfo
  {volume} {70}},\ \bibinfo {pages} {1787} (\bibinfo {year}
  {1993})}\BibitemShut {NoStop}%
\bibitem [{\citenamefont {Pikovsky}\ and\ \citenamefont
  {Shepelyansky}(2008)}]{Pikovsky2008}%
  \BibitemOpen
  \bibfield  {author} {\bibinfo {author} {\bibfnamefont {A.~S.}\ \bibnamefont
  {Pikovsky}}\ and\ \bibinfo {author} {\bibfnamefont {D.~L.}\ \bibnamefont
  {Shepelyansky}},\ }\href {\doibase 10.1103/PhysRevLett.100.094101} {\bibfield
   {journal} {\bibinfo  {journal} {Phys. Rev. Lett.}\ }\textbf {\bibinfo
  {volume} {100}},\ \bibinfo {pages} {094101} (\bibinfo {year}
  {2008})}\BibitemShut {NoStop}%
\bibitem [{\citenamefont {Flach}\ \emph {et~al.}(2009)\citenamefont {Flach},
  \citenamefont {Krimer},\ and\ \citenamefont {Skokos}}]{Flach2009}%
  \BibitemOpen
  \bibfield  {author} {\bibinfo {author} {\bibfnamefont {S.}~\bibnamefont
  {Flach}}, \bibinfo {author} {\bibfnamefont {D.~O.}\ \bibnamefont {Krimer}}, \
  and\ \bibinfo {author} {\bibfnamefont {C.}~\bibnamefont {Skokos}},\ }\href
  {\doibase 10.1103/PhysRevLett.102.024101} {\bibfield  {journal} {\bibinfo
  {journal} {Phys. Rev. Lett.}\ }\textbf {\bibinfo {volume} {102}},\ \bibinfo
  {pages} {024101} (\bibinfo {year} {2009})}\BibitemShut {NoStop}%
\bibitem [{\citenamefont {Gligori{\'{c} }}\ \emph {et~al.}(2011)\citenamefont
  {Gligori{\'{c} }}, \citenamefont {Bodyfelt},\ and\ \citenamefont
  {Flach}}]{Gligoric2011}%
  \BibitemOpen
  \bibfield  {author} {\bibinfo {author} {\bibfnamefont {G.}~\bibnamefont
  {Gligori{\'{c} }}}, \bibinfo {author} {\bibfnamefont {J.~D.}\ \bibnamefont
  {Bodyfelt}}, \ and\ \bibinfo {author} {\bibfnamefont {S.}~\bibnamefont
  {Flach}},\ }\href {\doibase 10.1209/0295-5075/96/30004} {\bibfield  {journal}
  {\bibinfo  {journal} {{EPL} (Europhysics Letters)}\ }\textbf {\bibinfo
  {volume} {96}},\ \bibinfo {pages} {30004} (\bibinfo {year}
  {2011})}\BibitemShut {NoStop}%
\bibitem [{\citenamefont {Cherroret}\ \emph {et~al.}(2014)\citenamefont
  {Cherroret}, \citenamefont {Vermersch}, \citenamefont {Garreau},\ and\
  \citenamefont {Delande}}]{Cherroret2014}%
  \BibitemOpen
  \bibfield  {author} {\bibinfo {author} {\bibfnamefont {N.}~\bibnamefont
  {Cherroret}}, \bibinfo {author} {\bibfnamefont {B.}~\bibnamefont
  {Vermersch}}, \bibinfo {author} {\bibfnamefont {J.~C.}\ \bibnamefont
  {Garreau}}, \ and\ \bibinfo {author} {\bibfnamefont {D.}~\bibnamefont
  {Delande}},\ }\href {\doibase 10.1103/PhysRevLett.112.170603} {\bibfield
  {journal} {\bibinfo  {journal} {Phys. Rev. Lett.}\ }\textbf {\bibinfo
  {volume} {112}},\ \bibinfo {pages} {170603} (\bibinfo {year}
  {2014})}\BibitemShut {NoStop}%
\bibitem [{\citenamefont {Lellouch}\ \emph {et~al.}(2020)\citenamefont
  {Lellouch}, \citenamefont {Ran{\c{c}}on}, \citenamefont {De~Bi\`evre},
  \citenamefont {Delande},\ and\ \citenamefont {Garreau}}]{Lellouch2020}%
  \BibitemOpen
  \bibfield  {author} {\bibinfo {author} {\bibfnamefont {S.}~\bibnamefont
  {Lellouch}}, \bibinfo {author} {\bibfnamefont {A.}~\bibnamefont
  {Ran{\c{c}}on}}, \bibinfo {author} {\bibfnamefont {S.}~\bibnamefont
  {De~Bi\`evre}}, \bibinfo {author} {\bibfnamefont {D.}~\bibnamefont
  {Delande}}, \ and\ \bibinfo {author} {\bibfnamefont {J.~C.}\ \bibnamefont
  {Garreau}},\ }\href {\doibase 10.1103/PhysRevA.101.043624} {\bibfield
  {journal} {\bibinfo  {journal} {Phys. Rev. A}\ }\textbf {\bibinfo {volume}
  {101}},\ \bibinfo {pages} {043624} (\bibinfo {year} {2020})}\BibitemShut
  {NoStop}%
\bibitem [{\citenamefont {Cazalilla}\ \emph {et~al.}(2011)\citenamefont
  {Cazalilla}, \citenamefont {Citro}, \citenamefont {Giamarchi}, \citenamefont
  {Orignac},\ and\ \citenamefont {Rigol}}]{Cazalilla2011}%
  \BibitemOpen
  \bibfield  {author} {\bibinfo {author} {\bibfnamefont {M.~A.}\ \bibnamefont
  {Cazalilla}}, \bibinfo {author} {\bibfnamefont {R.}~\bibnamefont {Citro}},
  \bibinfo {author} {\bibfnamefont {T.}~\bibnamefont {Giamarchi}}, \bibinfo
  {author} {\bibfnamefont {E.}~\bibnamefont {Orignac}}, \ and\ \bibinfo
  {author} {\bibfnamefont {M.}~\bibnamefont {Rigol}},\ }\href {\doibase
  10.1103/revmodphys.83.1405} {\bibfield  {journal} {\bibinfo  {journal} {Rev.
  Mod. Phys.}\ }\textbf {\bibinfo {volume} {83}},\ \bibinfo {pages} {1405}
  (\bibinfo {year} {2011})}\BibitemShut {NoStop}%
\bibitem [{\citenamefont {Cao}\ \emph {et~al.}(2022)\citenamefont {Cao},
  \citenamefont {Sajjad}, \citenamefont {Mas}, \citenamefont {Simmons},
  \citenamefont {Tanlimco}, \citenamefont {Nolasco-Martinez}, \citenamefont
  {Shimasaki}, \citenamefont {Kondakci}, \citenamefont {Galitski},\ and\
  \citenamefont {Weld}}]{Cao2022}%
  \BibitemOpen
  \bibfield  {author} {\bibinfo {author} {\bibfnamefont {A.}~\bibnamefont
  {Cao}}, \bibinfo {author} {\bibfnamefont {R.}~\bibnamefont {Sajjad}},
  \bibinfo {author} {\bibfnamefont {H.}~\bibnamefont {Mas}}, \bibinfo {author}
  {\bibfnamefont {E.~Q.}\ \bibnamefont {Simmons}}, \bibinfo {author}
  {\bibfnamefont {J.~L.}\ \bibnamefont {Tanlimco}}, \bibinfo {author}
  {\bibfnamefont {E.}~\bibnamefont {Nolasco-Martinez}}, \bibinfo {author}
  {\bibfnamefont {T.}~\bibnamefont {Shimasaki}}, \bibinfo {author}
  {\bibfnamefont {H.~E.}\ \bibnamefont {Kondakci}}, \bibinfo {author}
  {\bibfnamefont {V.}~\bibnamefont {Galitski}}, \ and\ \bibinfo {author}
  {\bibfnamefont {D.~M.}\ \bibnamefont {Weld}},\ }\href {\doibase
  10.1038/s41567-022-01724-7} {\bibfield  {journal} {\bibinfo  {journal}
  {Nature Physics}\ }\textbf {\bibinfo {volume} {18}},\ \bibinfo {pages} {1302}
  (\bibinfo {year} {2022})}\BibitemShut {NoStop}%
\bibitem [{\citenamefont {See~Toh}\ \emph {et~al.}(2022)\citenamefont
  {See~Toh}, \citenamefont {McCormick}, \citenamefont {Tang}, \citenamefont
  {Su}, \citenamefont {Luo}, \citenamefont {Zhang},\ and\ \citenamefont
  {Gupta}}]{SeeToh2022}%
  \BibitemOpen
  \bibfield  {author} {\bibinfo {author} {\bibfnamefont {J.~H.}\ \bibnamefont
  {See~Toh}}, \bibinfo {author} {\bibfnamefont {K.~C.}\ \bibnamefont
  {McCormick}}, \bibinfo {author} {\bibfnamefont {X.}~\bibnamefont {Tang}},
  \bibinfo {author} {\bibfnamefont {Y.}~\bibnamefont {Su}}, \bibinfo {author}
  {\bibfnamefont {X.-W.}\ \bibnamefont {Luo}}, \bibinfo {author} {\bibfnamefont
  {C.}~\bibnamefont {Zhang}}, \ and\ \bibinfo {author} {\bibfnamefont
  {S.}~\bibnamefont {Gupta}},\ }\href {\doibase 10.1038/s41567-022-01721-w}
  {\bibfield  {journal} {\bibinfo  {journal} {Nature Physics}\ }\textbf
  {\bibinfo {volume} {18}},\ \bibinfo {pages} {1297} (\bibinfo {year}
  {2022})}\BibitemShut {NoStop}%
\bibitem [{\citenamefont {Ermann}\ and\ \citenamefont
  {Shepelyansky}(2014)}]{Ermann2014}%
  \BibitemOpen
  \bibfield  {author} {\bibinfo {author} {\bibfnamefont {L.}~\bibnamefont
  {Ermann}}\ and\ \bibinfo {author} {\bibfnamefont {D.~L.}\ \bibnamefont
  {Shepelyansky}},\ }\href {\doibase 10.1088/1751-8113/47/33/335101} {\bibfield
   {journal} {\bibinfo  {journal} {Journal of Physics A: Mathematical and
  Theoretical}\ }\textbf {\bibinfo {volume} {47}},\ \bibinfo {pages} {335101}
  (\bibinfo {year} {2014})}\BibitemShut {NoStop}%
\bibitem [{\citenamefont {Vermersch}\ \emph {et~al.}(2020)\citenamefont
  {Vermersch}, \citenamefont {Delande},\ and\ \citenamefont
  {Garreau}}]{Vermersch2020}%
  \BibitemOpen
  \bibfield  {author} {\bibinfo {author} {\bibfnamefont {B.}~\bibnamefont
  {Vermersch}}, \bibinfo {author} {\bibfnamefont {D.}~\bibnamefont {Delande}},
  \ and\ \bibinfo {author} {\bibfnamefont {J.~C.}\ \bibnamefont {Garreau}},\
  }\href {\doibase 10.1103/PhysRevA.101.053625} {\bibfield  {journal} {\bibinfo
   {journal} {Phys. Rev. A}\ }\textbf {\bibinfo {volume} {101}},\ \bibinfo
  {pages} {053625} (\bibinfo {year} {2020})}\BibitemShut {NoStop}%
\bibitem [{\citenamefont {Slevin}\ and\ \citenamefont
  {Ohtsuki}(1999)}]{Slevin1999}%
  \BibitemOpen
  \bibfield  {author} {\bibinfo {author} {\bibfnamefont {K.}~\bibnamefont
  {Slevin}}\ and\ \bibinfo {author} {\bibfnamefont {T.}~\bibnamefont
  {Ohtsuki}},\ }\href {\doibase 10.1103/PhysRevLett.82.382} {\bibfield
  {journal} {\bibinfo  {journal} {Phys. Rev. Lett.}\ }\textbf {\bibinfo
  {volume} {82}},\ \bibinfo {pages} {382} (\bibinfo {year} {1999})}\BibitemShut
  {NoStop}%
\bibitem [{\citenamefont {Abrahams}\ \emph {et~al.}(1979)\citenamefont
  {Abrahams}, \citenamefont {Anderson}, \citenamefont {Licciardello},\ and\
  \citenamefont {Ramakrishnan}}]{Abrahams1979}%
  \BibitemOpen
  \bibfield  {author} {\bibinfo {author} {\bibfnamefont {E.}~\bibnamefont
  {Abrahams}}, \bibinfo {author} {\bibfnamefont {P.~W.}\ \bibnamefont
  {Anderson}}, \bibinfo {author} {\bibfnamefont {D.~C.}\ \bibnamefont
  {Licciardello}}, \ and\ \bibinfo {author} {\bibfnamefont {T.~V.}\
  \bibnamefont {Ramakrishnan}},\ }\href {\doibase 10.1103/PhysRevLett.42.673}
  {\bibfield  {journal} {\bibinfo  {journal} {Phys. Rev. Lett.}\ }\textbf
  {\bibinfo {volume} {42}},\ \bibinfo {pages} {673} (\bibinfo {year}
  {1979})}\BibitemShut {NoStop}%
\bibitem [{\citenamefont {W\"{o}lfle}\ and\ \citenamefont
  {Vollhardt}(2010)}]{Wolfle2010}%
  \BibitemOpen
  \bibfield  {author} {\bibinfo {author} {\bibfnamefont {P.}~\bibnamefont
  {W\"{o}lfle}}\ and\ \bibinfo {author} {\bibfnamefont {D.}~\bibnamefont
  {Vollhardt}},\ }\href {\doibase 10.1142/S0217979210064502} {\bibfield
  {journal} {\bibinfo  {journal} {International Journal of Modern Physics B}\
  }\textbf {\bibinfo {volume} {24}},\ \bibinfo {pages} {1526} (\bibinfo {year}
  {2010})}\BibitemShut {NoStop}%
\bibitem [{\citenamefont {Lopez}\ \emph {et~al.}(2013)\citenamefont {Lopez},
  \citenamefont {Clément}, \citenamefont {Lemarié}, \citenamefont {Delande},
  \citenamefont {Szriftgiser},\ and\ \citenamefont {Garreau}}]{Lopez2013}%
  \BibitemOpen
  \bibfield  {author} {\bibinfo {author} {\bibfnamefont {M.}~\bibnamefont
  {Lopez}}, \bibinfo {author} {\bibfnamefont {J.-F.}\ \bibnamefont {Clément}},
  \bibinfo {author} {\bibfnamefont {G.}~\bibnamefont {Lemarié}}, \bibinfo
  {author} {\bibfnamefont {D.}~\bibnamefont {Delande}}, \bibinfo {author}
  {\bibfnamefont {P.}~\bibnamefont {Szriftgiser}}, \ and\ \bibinfo {author}
  {\bibfnamefont {J.~C.}\ \bibnamefont {Garreau}},\ }\href {\doibase
  10.1088/1367-2630/15/6/065013} {\bibfield  {journal} {\bibinfo  {journal}
  {New Journal of Physics}\ }\textbf {\bibinfo {volume} {15}},\ \bibinfo
  {pages} {065013} (\bibinfo {year} {2013})}\BibitemShut {NoStop}%
\bibitem [{\citenamefont {Rodriguez}\ \emph {et~al.}(2011)\citenamefont
  {Rodriguez}, \citenamefont {Vasquez}, \citenamefont {Slevin},\ and\
  \citenamefont {R\"omer}}]{Rodriguez2011}%
  \BibitemOpen
  \bibfield  {author} {\bibinfo {author} {\bibfnamefont {A.}~\bibnamefont
  {Rodriguez}}, \bibinfo {author} {\bibfnamefont {L.~J.}\ \bibnamefont
  {Vasquez}}, \bibinfo {author} {\bibfnamefont {K.}~\bibnamefont {Slevin}}, \
  and\ \bibinfo {author} {\bibfnamefont {R.~A.}\ \bibnamefont {R\"omer}},\
  }\href {\doibase 10.1103/PhysRevB.84.134209} {\bibfield  {journal} {\bibinfo
  {journal} {Phys. Rev. B}\ }\textbf {\bibinfo {volume} {84}},\ \bibinfo
  {pages} {134209} (\bibinfo {year} {2011})}\BibitemShut {NoStop}%
\bibitem [{\citenamefont {Akridas-Morel}\ \emph {et~al.}(2019)\citenamefont
  {Akridas-Morel}, \citenamefont {Cherroret},\ and\ \citenamefont
  {Delande}}]{Panayotis2019}%
  \BibitemOpen
  \bibfield  {author} {\bibinfo {author} {\bibfnamefont {P.}~\bibnamefont
  {Akridas-Morel}}, \bibinfo {author} {\bibfnamefont {N.}~\bibnamefont
  {Cherroret}}, \ and\ \bibinfo {author} {\bibfnamefont {D.}~\bibnamefont
  {Delande}},\ }\href {\doibase 10.1103/PhysRevA.100.043612} {\bibfield
  {journal} {\bibinfo  {journal} {Phys. Rev. A}\ }\textbf {\bibinfo {volume}
  {100}},\ \bibinfo {pages} {043612} (\bibinfo {year} {2019})}\BibitemShut
  {NoStop}%
\bibitem [{\citenamefont {Chalker}(1990)}]{Chalker1990}%
  \BibitemOpen
  \bibfield  {author} {\bibinfo {author} {\bibfnamefont {J.}~\bibnamefont
  {Chalker}},\ }\href {\doibase https://doi.org/10.1016/0378-4371(90)90056-X}
  {\bibfield  {journal} {\bibinfo  {journal} {Physica A: Statistical Mechanics
  and its Applications}\ }\textbf {\bibinfo {volume} {167}},\ \bibinfo {pages}
  {253} (\bibinfo {year} {1990})}\BibitemShut {NoStop}%
\bibitem [{\citenamefont {Girardeau}(1960)}]{Girardeau1960}%
  \BibitemOpen
  \bibfield  {author} {\bibinfo {author} {\bibfnamefont {M.}~\bibnamefont
  {Girardeau}},\ }\href {\doibase 10.1063/1.1703687} {\bibfield  {journal}
  {\bibinfo  {journal} {Journal of Mathematical Physics}\ }\textbf {\bibinfo
  {volume} {1}},\ \bibinfo {pages} {516} (\bibinfo {year} {1960})}\BibitemShut
  {NoStop}%
\bibitem [{\citenamefont {Lenard}(1964)}]{Lenard1964}%
  \BibitemOpen
  \bibfield  {author} {\bibinfo {author} {\bibfnamefont {A.}~\bibnamefont
  {Lenard}},\ }\href {\doibase 10.1063/1.1704196} {\bibfield  {journal}
  {\bibinfo  {journal} {Journal of Mathematical Physics}\ }\textbf {\bibinfo
  {volume} {5}},\ \bibinfo {pages} {930} (\bibinfo {year} {1964})}\BibitemShut
  {NoStop}%
\bibitem [{\citenamefont {Buljan}\ \emph {et~al.}(2008)\citenamefont {Buljan},
  \citenamefont {Pezer},\ and\ \citenamefont {Gasenzer}}]{Buljan2008}%
  \BibitemOpen
  \bibfield  {author} {\bibinfo {author} {\bibfnamefont {H.}~\bibnamefont
  {Buljan}}, \bibinfo {author} {\bibfnamefont {R.}~\bibnamefont {Pezer}}, \
  and\ \bibinfo {author} {\bibfnamefont {T.}~\bibnamefont {Gasenzer}},\ }\href
  {\doibase 10.1103/PhysRevLett.100.080406} {\bibfield  {journal} {\bibinfo
  {journal} {Phys. Rev. Lett.}\ }\textbf {\bibinfo {volume} {100}},\ \bibinfo
  {pages} {080406} (\bibinfo {year} {2008})}\BibitemShut {NoStop}%
\bibitem [{\citenamefont {Juki\ifmmode~\acute{c}\else \'{c}\fi{}}\ \emph
  {et~al.}(2008)\citenamefont {Juki\ifmmode~\acute{c}\else \'{c}\fi{}},
  \citenamefont {Pezer}, \citenamefont {Gasenzer},\ and\ \citenamefont
  {Buljan}}]{Jukic2008}%
  \BibitemOpen
  \bibfield  {author} {\bibinfo {author} {\bibfnamefont {D.}~\bibnamefont
  {Juki\ifmmode~\acute{c}\else \'{c}\fi{}}}, \bibinfo {author} {\bibfnamefont
  {R.}~\bibnamefont {Pezer}}, \bibinfo {author} {\bibfnamefont
  {T.}~\bibnamefont {Gasenzer}}, \ and\ \bibinfo {author} {\bibfnamefont
  {H.}~\bibnamefont {Buljan}},\ }\href {\doibase 10.1103/PhysRevA.78.053602}
  {\bibfield  {journal} {\bibinfo  {journal} {Phys. Rev. A}\ }\textbf {\bibinfo
  {volume} {78}},\ \bibinfo {pages} {053602} (\bibinfo {year}
  {2008})}\BibitemShut {NoStop}%
\bibitem [{\citenamefont {Pezer}\ \emph {et~al.}(2009)\citenamefont {Pezer},
  \citenamefont {Gasenzer},\ and\ \citenamefont {Buljan}}]{Pezer2009}%
  \BibitemOpen
  \bibfield  {author} {\bibinfo {author} {\bibfnamefont {R.}~\bibnamefont
  {Pezer}}, \bibinfo {author} {\bibfnamefont {T.}~\bibnamefont {Gasenzer}}, \
  and\ \bibinfo {author} {\bibfnamefont {H.}~\bibnamefont {Buljan}},\ }\href
  {\doibase 10.1103/PhysRevA.80.053616} {\bibfield  {journal} {\bibinfo
  {journal} {Phys. Rev. A}\ }\textbf {\bibinfo {volume} {80}},\ \bibinfo
  {pages} {053616} (\bibinfo {year} {2009})}\BibitemShut {NoStop}%
\bibitem [{\citenamefont {Rigol}\ and\ \citenamefont
  {Muramatsu}(2005{\natexlab{a}})}]{Rigol2005a}%
  \BibitemOpen
  \bibfield  {author} {\bibinfo {author} {\bibfnamefont {M.}~\bibnamefont
  {Rigol}}\ and\ \bibinfo {author} {\bibfnamefont {A.}~\bibnamefont
  {Muramatsu}},\ }\href {\doibase 10.1103/PhysRevLett.94.240403} {\bibfield
  {journal} {\bibinfo  {journal} {Phys. Rev. Lett.}\ }\textbf {\bibinfo
  {volume} {94}},\ \bibinfo {pages} {240403} (\bibinfo {year}
  {2005}{\natexlab{a}})}\BibitemShut {NoStop}%
\bibitem [{\citenamefont {Rigol}\ and\ \citenamefont
  {Muramatsu}(2005{\natexlab{b}})}]{Rigol2005}%
  \BibitemOpen
  \bibfield  {author} {\bibinfo {author} {\bibfnamefont {M.}~\bibnamefont
  {Rigol}}\ and\ \bibinfo {author} {\bibfnamefont {A.}~\bibnamefont
  {Muramatsu}},\ }\href {\doibase 10.1103/PhysRevA.72.013604} {\bibfield
  {journal} {\bibinfo  {journal} {Phys. Rev. A}\ }\textbf {\bibinfo {volume}
  {72}},\ \bibinfo {pages} {013604} (\bibinfo {year}
  {2005}{\natexlab{b}})}\BibitemShut {NoStop}%
\bibitem [{\citenamefont {Olshanii}\ and\ \citenamefont
  {Dunjko}(2003)}]{Olshanii2003}%
  \BibitemOpen
  \bibfield  {author} {\bibinfo {author} {\bibfnamefont {M.}~\bibnamefont
  {Olshanii}}\ and\ \bibinfo {author} {\bibfnamefont {V.}~\bibnamefont
  {Dunjko}},\ }\href {\doibase 10.1103/physrevlett.91.090401} {\bibfield
  {journal} {\bibinfo  {journal} {Phys. Rev. Lett.}\ }\textbf {\bibinfo
  {volume} {91}},\ \bibinfo {pages} {090401} (\bibinfo {year}
  {2003})}\BibitemShut {NoStop}%
\bibitem [{\citenamefont {Tan}(2008)}]{Tan2008}%
  \BibitemOpen
  \bibfield  {author} {\bibinfo {author} {\bibfnamefont {S.}~\bibnamefont
  {Tan}},\ }\href {\doibase https://doi.org/10.1016/j.aop.2008.03.005}
  {\bibfield  {journal} {\bibinfo  {journal} {Annals of Physics}\ }\textbf
  {\bibinfo {volume} {323}},\ \bibinfo {pages} {2971} (\bibinfo {year}
  {2008})}\BibitemShut {NoStop}%
\bibitem [{\citenamefont {Vignolo}\ and\ \citenamefont
  {Minguzzi}(2013)}]{Vignolo2013}%
  \BibitemOpen
  \bibfield  {author} {\bibinfo {author} {\bibfnamefont {P.}~\bibnamefont
  {Vignolo}}\ and\ \bibinfo {author} {\bibfnamefont {A.}~\bibnamefont
  {Minguzzi}},\ }\href {\doibase 10.1103/physrevlett.110.020403} {\bibfield
  {journal} {\bibinfo  {journal} {Phys. Rev. Lett.}\ }\textbf {\bibinfo
  {volume} {110}},\ \bibinfo {pages} {020403} (\bibinfo {year}
  {2013})}\BibitemShut {NoStop}%
\bibitem [{\citenamefont {Its}\ \emph {et~al.}(1991)\citenamefont {Its},
  \citenamefont {Izergin},\ and\ \citenamefont {Korepin}}]{Its1991}%
  \BibitemOpen
  \bibfield  {author} {\bibinfo {author} {\bibfnamefont {A.}~\bibnamefont
  {Its}}, \bibinfo {author} {\bibfnamefont {A.}~\bibnamefont {Izergin}}, \ and\
  \bibinfo {author} {\bibfnamefont {V.}~\bibnamefont {Korepin}},\ }\href
  {\doibase https://doi.org/10.1016/0167-2789(91)90171-5} {\bibfield  {journal}
  {\bibinfo  {journal} {Physica D: Nonlinear Phenomena}\ }\textbf {\bibinfo
  {volume} {53}},\ \bibinfo {pages} {187} (\bibinfo {year} {1991})}\BibitemShut
  {NoStop}%
\bibitem [{\citenamefont {Rigol}(2005)}]{Rigol2005temp}%
  \BibitemOpen
  \bibfield  {author} {\bibinfo {author} {\bibfnamefont {M.}~\bibnamefont
  {Rigol}},\ }\href {\doibase 10.1103/PhysRevA.72.063607} {\bibfield  {journal}
  {\bibinfo  {journal} {Phys. Rev. A}\ }\textbf {\bibinfo {volume} {72}},\
  \bibinfo {pages} {063607} (\bibinfo {year} {2005})}\BibitemShut {NoStop}%
\end{thebibliography}%



\end{document}